\documentclass[aps, pra, groupedaddress, twocolumn, floatfix, 10pt]{revtex4-2}
\usepackage{amssymb}
\usepackage{amsmath}
\usepackage{graphicx}
\usepackage[caption=false, subrefformat=parens,labelformat=parens]{subfig}
\usepackage{dcolumn}
\usepackage{booktabs}
\usepackage{longtable}
\usepackage{xcolor}
\usepackage{hyperref}
\newcommand*\mean[1]{\overline{#1}}

\begin{document}

\title{A New Attempt to Identify Long-term Precursors for Endogenous Financial
	Crises\texorpdfstring{\\ in the Market Correlation Structures }{}}

\author{Anton J. Heckens}
\email{anton.heckens@uni-due.de}
\author{Thomas Guhr}
\email{thomas.guhr@uni-due.de}
\affiliation{
	Fakult\"at f\"ur Physik, Universit\"at Duisburg-Essen, Duisburg, Germany		
}

\begin{abstract}
  Prediction of events in financial markets is every investor's dream
  and, usually, wishful thinking. From a more general, economic and
  societal viewpoint, the identification of indicators for large
  events is highly desirable to assess systemic risks. Unfortunately,
  the very nature of financial markets, particularly the predominantly
  non-Markovian character as well as non-stationarity, make this
  challenge a formidable one, leaving little hope for fully fledged
  answers. Nevertheless, it is called for to collect pieces of
  evidence in a variety of observables to be assembled like the pieces
  of a puzzle that eventually might help to catch a glimpse of long-term
  indicators or precursors for large events -- if at all in a
  statistical sense. Here, we present a new piece for this puzzle.  We
  use the quasi-stationary market states which exist in the time
  evolution of the correlation structure in financial
  markets. Recently, we identified such market states relative to the
  collective motion of the market as a whole.  We study their
  precursor properties in the US stock markets over 16 years,
  including two endogenous crises, the dot-com bubble burst and the pre-phase
  of the Lehman Brothers crash.  We identify certain interesting features
  and critically discuss their suitability as indicators.
\end{abstract}

\maketitle

\section{\label{sec:Introduction}Introduction}

Critical events in the financial markets bear, in the age of
globalization, ever higher risks for the world's economic system,
which then, in a feedback loop, can cause negative impacts on the
financial markets. An early warning system is as desirable as in the
case of geologic, seismic and volcanic hazards, but at least as
difficult to design. The natural laws under which the latter emerge do
not alter, while the fast economic and societal development implies,
for the structure and the functionality of the financial system, a new
and considerable unpredictability which adds to the risks due to the
kind of non-stationarity which has always been present in the
markets~\cite{Mandelbrot1997,schwert1989does,LONGIN19953,mantegna1999introduction,bouchaud2003theory,Kutner_2019}.

Musmeci et al.~\cite{musmeci2016interplay} showed that the correlation
structure can be used, to some extent, to forecast the volatilities
within ``persistent'' periods, \textit{i.e.}~periods of
quasi-stationary correlation structures.  This ends when transitions
between persistent periods take place which is often followed by
larger volatility changes.  These persistent periods are strongly
connected to ``market states'' or ``regimes'' in the economics
terminology~\cite{campbellEconometricsFinancialMarkets1997,
  munnixIdentifyingStatesFinancial2012,Hamilton_1989,Hamilton_1990,
  Schaller_1997,Matteo_2002,jurczyk2017measuring}.  There are different ways to define
market states~\cite{marti2021review}. Here, we follow
Refs.~\cite{munnixIdentifyingStatesFinancial2012,Stepanov_2015,Rinn_2015,Chetalova_2015,Chetalova_2015_2,Heckens_2020,Wang_2020}
and identify quasi-stationary markets
states in the time evolution of the non-stationary correlation
structure. The industrial sectors, which are clearly visible in the
correlations~\cite{Laloux_1999, Noh_2000, Gopikrishnan_2001,
  Plerou_2002, Borghesi_2007, kenett2009rmt, shapira2009index,
  MacMahon_2015} and covariances~\cite{Benzaquen_2017}, and their mutual connections are thereby analyzed
in a time resolved fashion.  This is
accomplished by applying $k$-means clustering~\cite{Steinhaus1956:DivisionCorpMaterielEnParties,
	BallHall1965:ISODATA, MacQueen1967:SomeMethodsClassification,
	LLoyd1982:LeastSquaresQuantization, Jain_2010}, a machine learning
algorithm, to a set of correlation matrices measured over time in a
moving window. The resulting
clusters which we identify as market states can be regarded as quasi-stationary structures in time, such that 
the individual
correlation matrices fluctuate about the cluster centers~\cite{Stepanov_2015,Chetalova_2015}.
The market states emerge, exist for some time and eventually
disappear~\cite{munnixIdentifyingStatesFinancial2012,Stepanov_2015,Rinn_2015,Chetalova_2015,Chetalova_2015_2,
  papenbrockHandlingRiskonRiskoff2015,Qiu_2018,Heckens_2020,pharasi2020market,
  pharasi2020dynamics,Wang_2020,pharasi2021dynamics}.  Recently, similar methods have been
applied in other fields, such as studies of epileptic
seizures~\cite{rings2019traceability} of freeway traffic~\cite{Wang_2020}.

The market states are known to be dominated by the collective motion
of the market as a whole~\cite{Stepanov_2015} which is related to the
fact that the corresponding eigenvalue is largest and a measure for
the average correlation coefficient~\cite{Song_2011}.  To uncover the dynamics of the correlation
structure, particularly due to the industrial sectors, relative to the
dominating collective motion, \textit{i.e.}~to measure the
correlations in the moving frame of the collective motion, we recently
put forward a systematic and mathematically well-defined
method~\cite{Heckens_2020}.  We subtract the dyadic matrix belonging
to the largest eigenvalue and thereby define two types of
reduced-rank correlation matrices whose dynamics are then analyzed
with the above  $k$-means clustering algorithm.
Reduced-rank correlation matrices of another kind appear in the context of filtering where the smaller eigenvalues are removed as a noise reduction technique~\cite{Laloux_1999,Alter_2000,Kim_2005,MacMahon_2015,laloux2000random}.
Subsequently, from the filtered standard correlation matrices, reduced-rank correlation matrices are calculated in order to obtain well-defined correlation matrices \cite{Miceli_2004,
Tumminello_2007, tumminello2007shrinkage}.
As exogenous effects often affect the market as a whole, our reduced-rank correlation matrices are likely to be dominated by endogenous effects~\cite{Heckens_2020}.

The quasi-stationary market states in the time evolution of these
reduced-rank correlation matrices are not directly related to the
ones of the standard, \textit{i.e.}~not reduced-rank,
ones. Surprisingly, they appear to be more stable, sometimes over
several years, than the ones for the standard correlation matrices.
This observation prompted the present study.  Pharasi et
al.~\cite{Pharasi_2018} proposed to identify precursors for critical
events in the quasi-stationary market states of the standard
correlation matrices, while we here use the reduced-rank ones in view
of their higher stability. More precisely, we exploit the separation
of two different time scales. The quickly changing collective
contribution is separated from the more stable one due to the
industrial sectors. 

Furthermore, we investigate the average dynamics of standard and reduced-rank matrices 
taking the mean value of all matrix elements. 
This adds to preceding studies on volatilities such as the absolute value of returns~\cite{DING_1993}
or cross-sectional volatilities and correlations~\cite{Campbell_1999,Solnik_2000,Ankrim_2002}.

It is worth mentioning that the collective market
motion is at least more likely to be influenced by exogenous effects
than the motion relative to it. Exogenous information can trigger
collective volatility outbursts as the traders' reaction to unexpected
events~\cite{sornette2003causes, bouchaud2010endogenous,
  Danielsson_2012, Bouchaud_2011}. By removing these risk
contributions, we analyze the dynamics of the endogenous risk in the
mutual interaction of the industrial sectors and the corresponding
contagious effect for the whole market.
For instance, we search for drastic changes in the
financial sector in the pre-phase of the Lehman crash as an endogenous crisis which emerged from a major crisis in the US housing market.
Such changes to which we refer as precursors (cf.~Ref.~\cite{Pharasi_2018}) indicate systemic instabilities not necessarily but potentially leading to a crash.
Our analysis adds to previous studies of systemic risk~\cite{Bisias_2012}.
Noteworthy are investigations related to principal components analysis~\cite{zheng2012changes,Billio_2012,kritzman2011principal} and causality measures~\cite{Billio_2012,beguvsic2018information}.

Here, we take data into account exclusively from epochs before
crises events or within crises periods in order to investigate market
state transitions as precursors.  We address these two questions: How
many epochs before a crisis event or within a crisis period does a
``crisis market state'' in the correlation structure show up? -- Can
we identify characteristic precursor signals in the dynamics of the
reduced-ranked correlation matrices without using cluster methods?

The paper is organized as follows.  In
Sec.~\ref{sec:DataSet}, we introduce our set of daily
data for the analysis. The construction of reduced-rank correlation
matrices is briefly sketched in
Sec.~\ref{sec:ConstructionOfReducedRankMatrices}. In
Sec.~\ref{sec:DataAnalysisAndResults}, we analyze the data and present our results.
The conclusions are given in Sec.~\ref{sec:Conclusion}.

\section{\label{sec:DataSet}Data set}

Using data collected by QuoteMedia~\cite{DynamicStockMarket} and provided by Quandl \cite{Quandl} we construct a data set
of $K=250$ US stocks (see~Tab.~\ref{tab:OverviewSP500} in App.~\ref{sec:ListStocks}), \textit{i.e.} the selected stocks do not change for the entire investigation period.
The investigation period ranges from 02~January,~1997 to 31~December,~2012 in order to concentrate on the two
periods, the dot-com bubble burst and the Lehman Brothers pre-crash phase.
Our data set represents the S\&P\,500 index since it comprises the 11 Global Industry Classification Standard (GICS) sectors in Tab.~\ref{tab:GICS} (cf.~\cite{wiki:2020:GICS}).
Furthermore, we also sorted the stocks within the sectors according to the sub-industry sectors (see~Tab.~\ref{tab:OverviewSP500} in App.~\ref{sec:ListStocks}).
\begin{table}[!htb]
	\centering
	\caption{\label{tab:GICS}
		Global Industry Classification Standard (GICS).
	}
	\begin{tabular}{l@{\hspace{5em}}l@{\hspace{2em}}d}
		\toprule
		\multicolumn{1}{l}{Abbreviation} &
		\multicolumn{1}{l}{Sector}{}  &
		\multicolumn{1}{c}{Number of} \\
		\multicolumn{1}{l}{} &
		\multicolumn{1}{l}{} &
		\multicolumn{1}{c}{stocks}
		\\
		\midrule
		E & Energy & 16 \\
		M & Materials & 14 \\
		I & Industrials & 43 \\
		CD & Consumer Discretionary & 24 \\
		CST & Consumer Staples & 24 \\
		HC & Health Care & 28 \\
		F & Financials & 36 \\
		RE & Real Estate & 8 \\
		I & Information Technology & 27 \\
		CSE & Communication Services & 9 \\
		U & Utilities & 21 \\
		\bottomrule
	\end{tabular}
\end{table}
In Ref.~\cite{Heckens_2020}, we demonstrated that the market states depend on the choice of the stocks. Nevertheless, our data set allows us to make general statements about the correlation structure of the US stock markets
because all GICS industry sectors \cite{wiki:2020:GICS} are covered by our data set and no industry sector is underrepresented, even not the real estate sector which mainly causes
a market state transition in Ref.~\cite{Heckens_2020}.

From adjusted closing prices $S_i(t)$, we calculate the daily logarithmic returns ($\Delta t = 1$ day)
\begin{equation} \label{eqn:LogReturn}
	G_i(t) =  \ln  \frac{S_i(t+\Delta t)}{S_i(t)} , \hspace{0.5cm} i = 1, 
	\ldots, K
	\,.
\end{equation}
We set up a $K \times T_{\text{tot}}$ data matrix for the total investigation period
\begin{equation} \label{eqn:DatamatrixG}
	G_{\text{tot}} = \begin{bmatrix}  G_1(1) & \dots & G_1(T_{\text{tot}})  \\
		\vdots & & \vdots \\
		G_i(1) & \dots & G_i(T_{\text{tot}}) \\
		\vdots & & \vdots \\
		G_{K}(1) & \dots & G_K(T_{\text{tot}}) 
	\end{bmatrix}
\end{equation}
with $K= 250$, being the number of stocks and $T_\text{tot} = 4026$,
being the total number of points in the return time series of a stock. 
We do not use the full data matrix $G_{\text{tot}}$ for the market state analysis. In order to analyze the non-stationarity of correlation matrices we select subblocks of the data matrix $G_{\text{tot}}$ with all $K= 250$ stocks and intervals of $T_\text{ep} = 42$ trading days 
which correspond to 2 trading months.
Only in the case of disjoint intervals we refer to these intervals as epochs.

\section{\label{sec:ConstructionOfReducedRankMatrices}Reduced-Rank Matrices}

\begin{figure*}[!htb]
	% \vspace{-1cm}
	\centering
	\subfloat[\label{subfig:CorrMatEntTimePeriodCovAppr}Covariance approach.]{
		\includegraphics[width=0.5\textwidth]{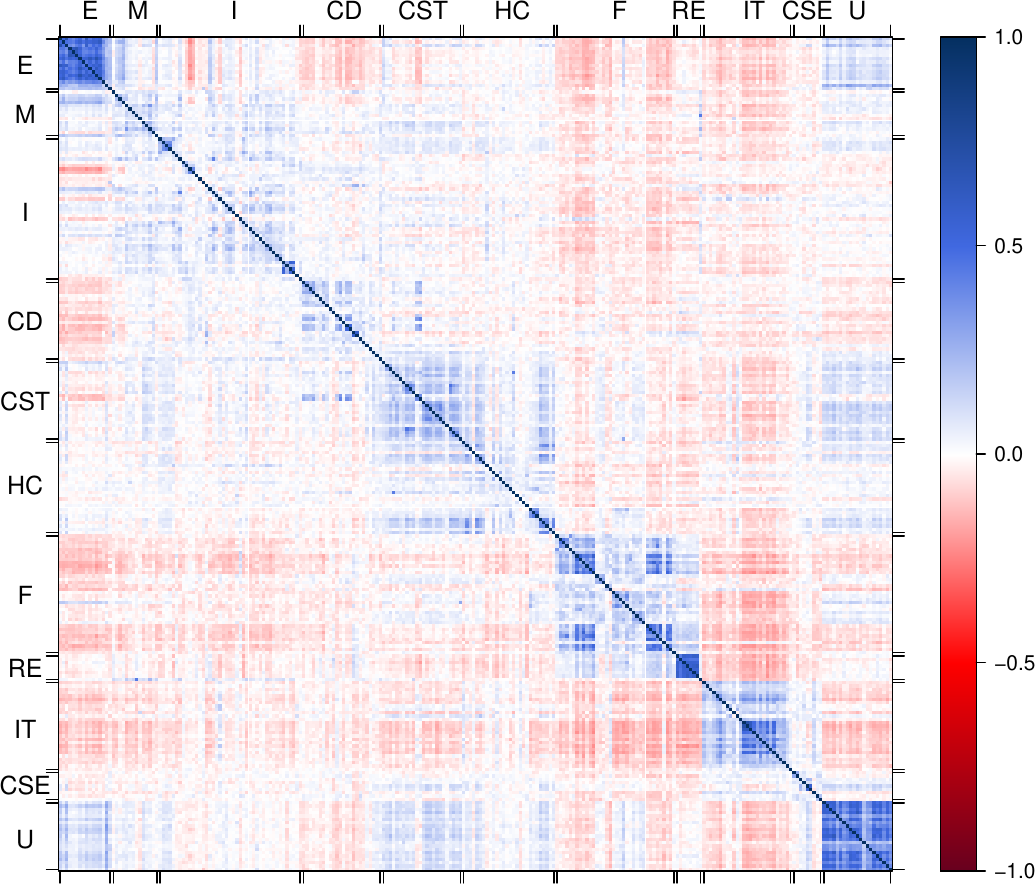}
	} %\\
	\subfloat[\label{subfig:CorrMatEntTimePeriodCorrAppr}Correlation approach.]{
		\includegraphics[width=0.5\textwidth]{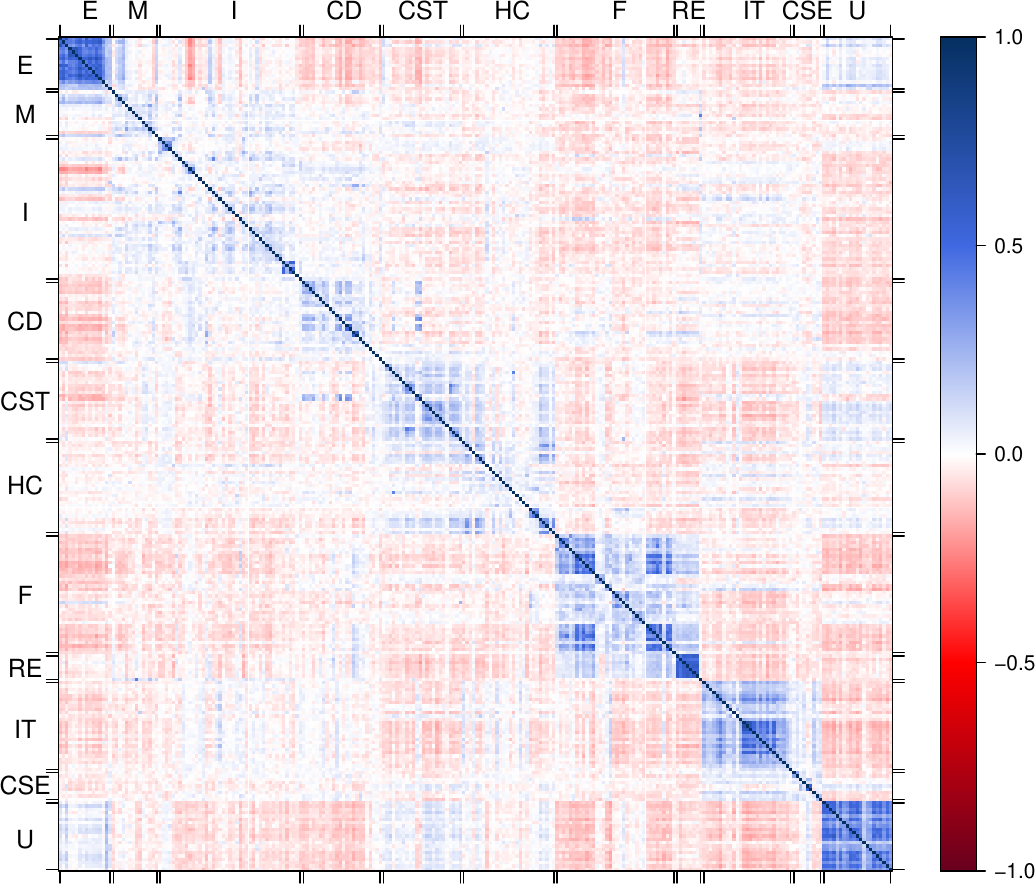}
	}
	\caption{\label{subfig:Main:CorrMatEntTimePeriod} Reduced-rank correlation matrices of $K = 250$ stocks for \protect\subref{subfig:CorrMatEntTimePeriodCovAppr} the covariance approach and \protect\subref{subfig:CorrMatEntTimePeriodCorrAppr} the correlation approach. Both matrices are calculated for the 16 year period from 02~January,~1997 to 31~December,~2012. Capital Letters indicate industrial sectors (see~Tab.~\ref{tab:GICS}) (\href{https://www.quandl.com/}{Data from QuoteMedia via Quandl}).}
\end{figure*}
\begin{figure*}[!htb]
	% \vspace{-1cm}
	\centering
	\begin{minipage}{0.5\textwidth}
		\subfloat[\label{subfig:CovStandFraction}]
		{\includegraphics[width=1.0\linewidth]{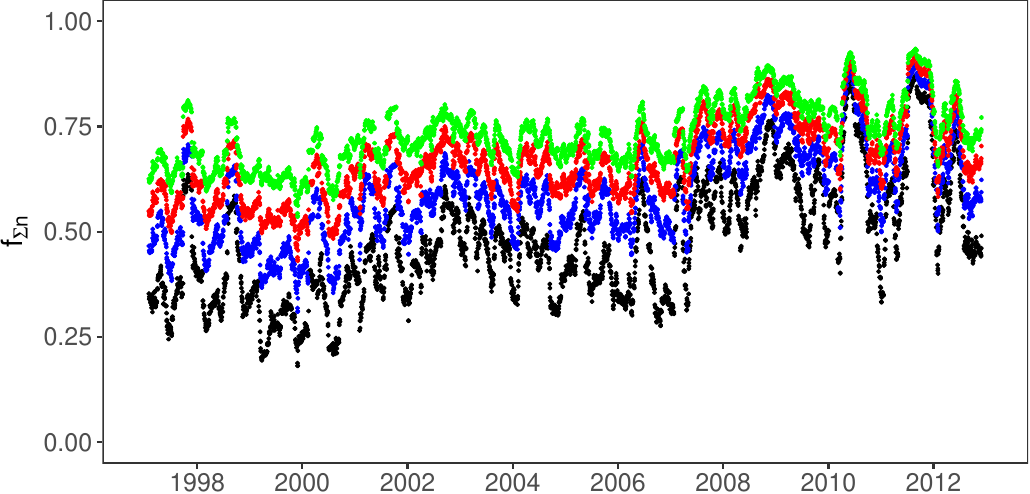}
		}\par
		\subfloat[\label{subfig:CovSubstrFraction}]
		{\includegraphics[width=1.0\textwidth]{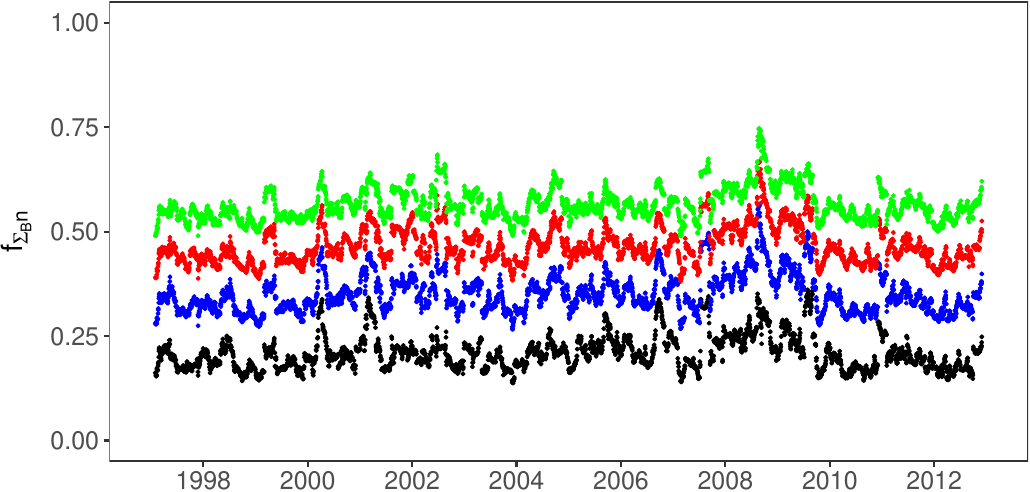}
		}\par
		\subfloat[\label{subfig:CorrSubstrFraction}]
		{\includegraphics[width=1.0\textwidth]{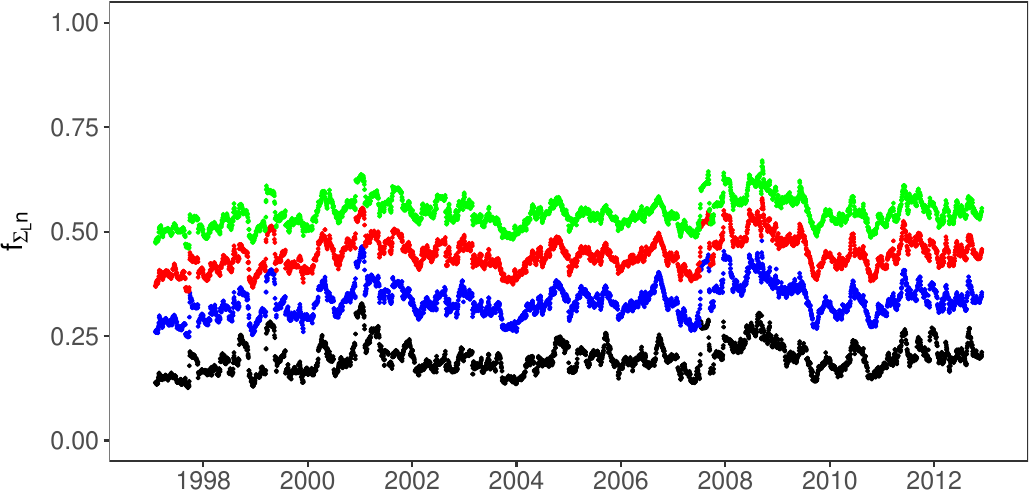}
		}%
	\end{minipage}%
	\begin{minipage}{0.5\textwidth}
		\subfloat[\label{subfig:CorrStandFraction}]
		{\includegraphics[width=1.0\textwidth]{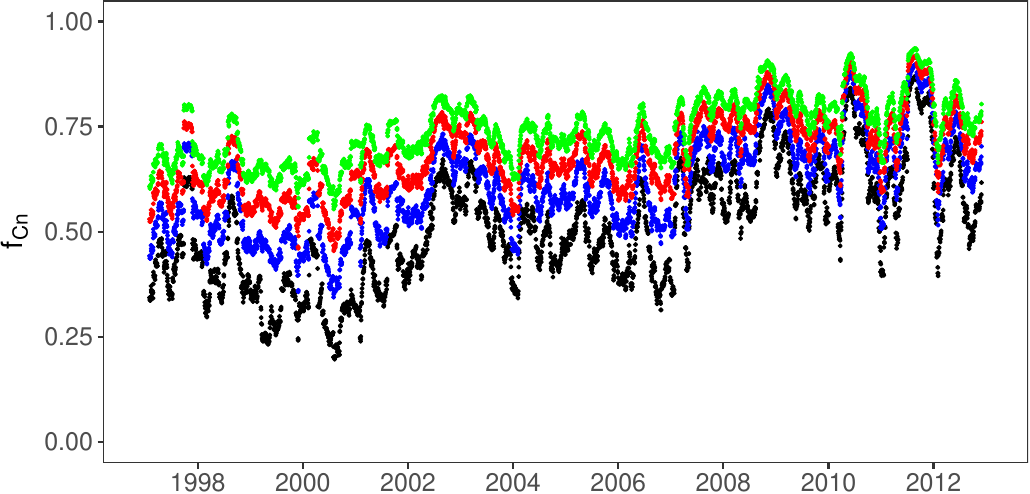}
		}\par
		\subfloat[\label{subfig:CovRedRankFraction}]
		{\includegraphics[width=1.0\textwidth]{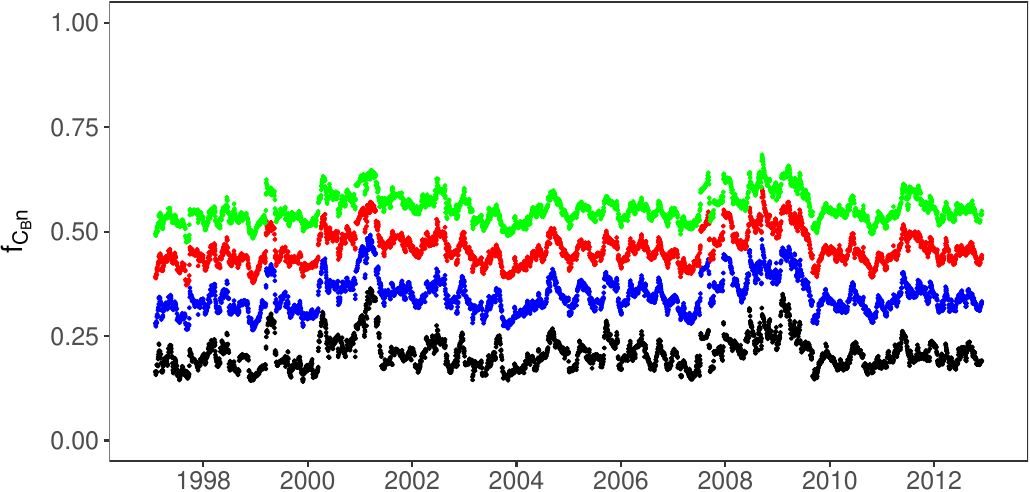}
		}\par
		\subfloat[\label{subfig:CorrRedRankFraction}]
		{\includegraphics[width=1.0\textwidth]{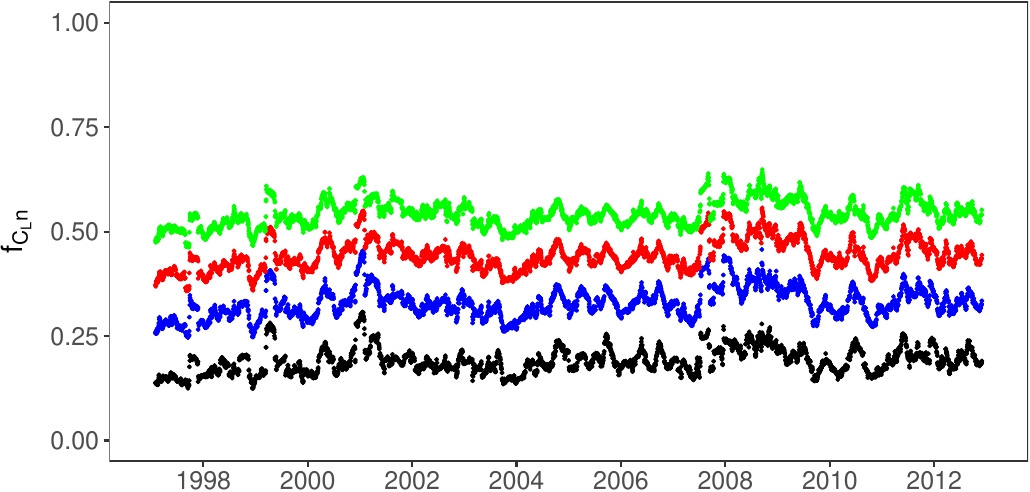}
		}
	\end{minipage}
	\caption{\label{subfig:Main:Fractions} Comparison of different fractions: \protect\subref{subfig:CovStandFraction} $f_{\Sigma n}$, \protect\subref{subfig:CovSubstrFraction} $f_{\Sigma_B n}$,
		\protect\subref{subfig:CorrSubstrFraction} $f_{\Sigma_L n}$,
		\protect\subref{subfig:CorrStandFraction} $f_{C n}$,  
		\protect\subref{subfig:CovRedRankFraction} $f_{C_B n}$ and \protect\subref{subfig:CorrRedRankFraction} $f_{C_L n}$ for $n=1$ (black), $n=2$ (blue), $n=3$ (red), $n=4$ (green).
		Each dot represents an interval of 42 trading days (\href{https://www.quandl.com/}{Data 
			from QuoteMedia via Quandl}).}
\end{figure*}
\begin{figure*}[!htb]
	% \vspace{-1cm}
	\centering
	\begin{minipage}{0.5\textwidth}
		\subfloat[\label{subfig:CovStandMean}Mean covariance $\mean{\text{cov}}$.]
		{\includegraphics[width=1.0\linewidth]{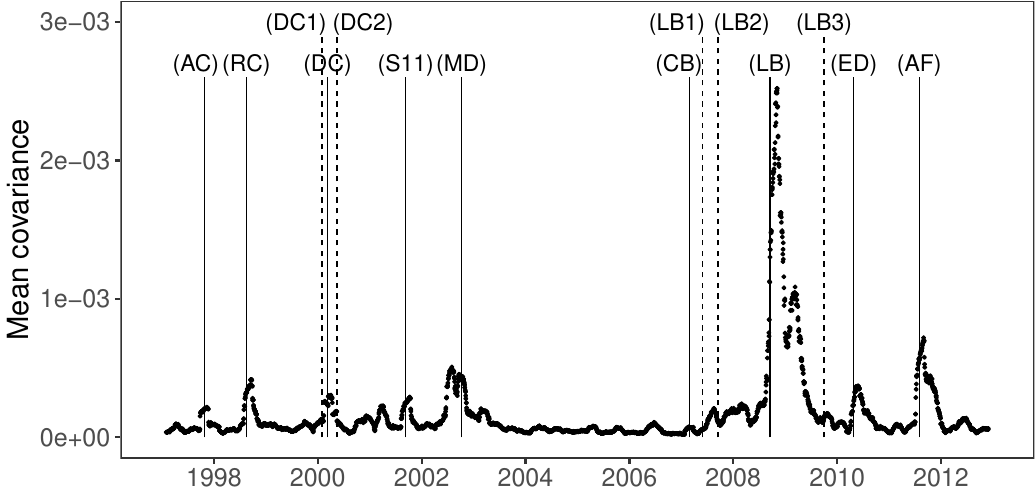}
		}\par
		\subfloat[\label{subfig:CovSubstrDyadMean}Mean covariance $\mean{\text{cov}}_B$.]
		{\includegraphics[width=1.0\textwidth]{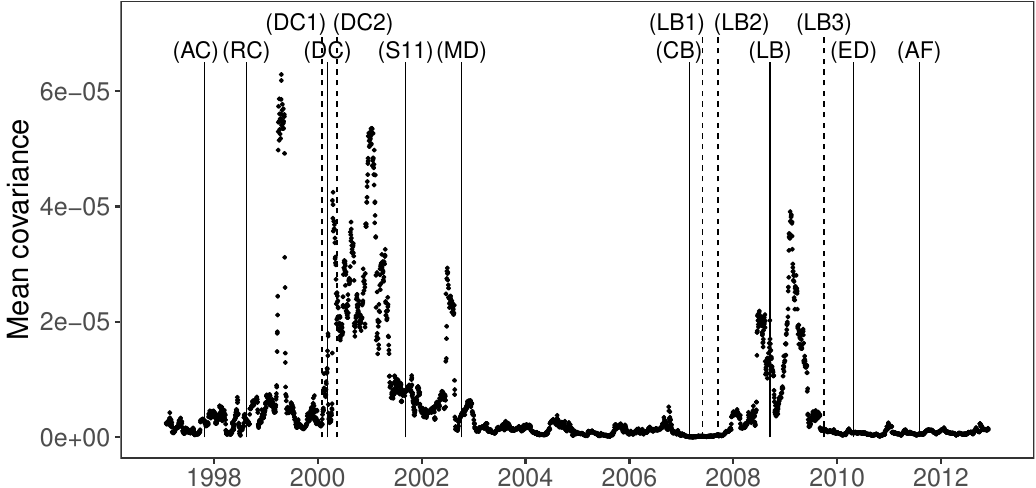}
		}\par
		\subfloat[\label{subfig:CorrSubstrDyadMean}Mean covariance $\mean{\text{cov}}_L$.]
		{\includegraphics[width=1.0\textwidth]{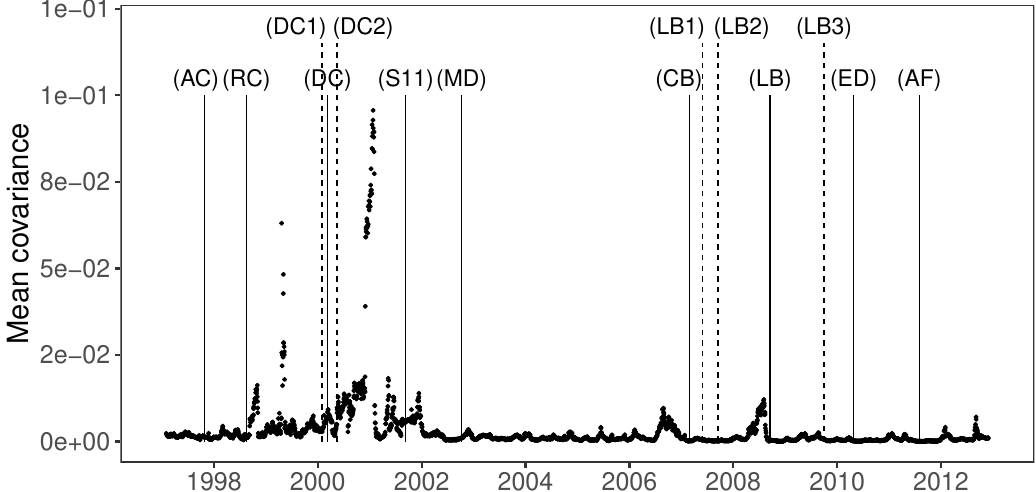}
		}%
	\end{minipage}%
	\begin{minipage}{0.5\textwidth}
		\subfloat[\label{subfig:CorrStandMean}Mean correlation $\mean{\text{corr}}$.]
		{\includegraphics[width=1.0\textwidth]{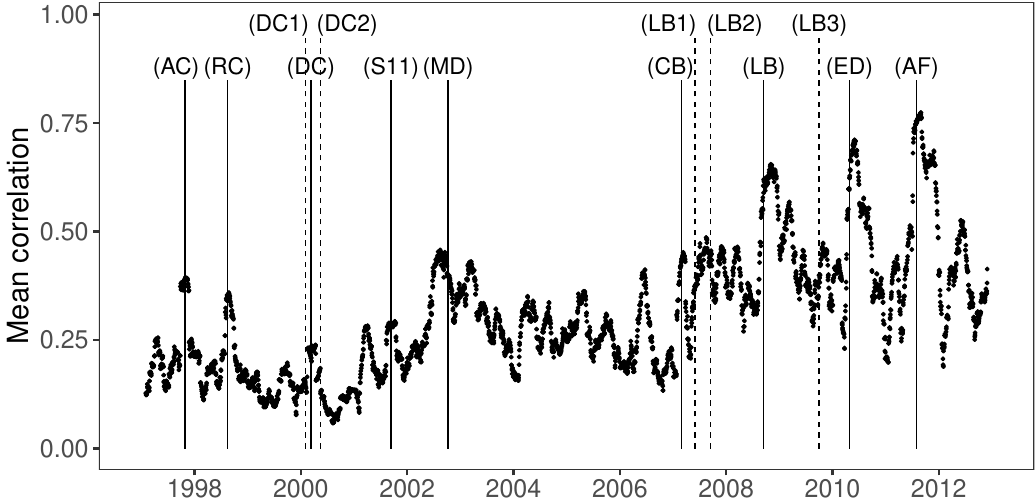}
		}\par
		\subfloat[\label{subfig:CovRedRankMean}Mean correlation $\mean{\text{corr}}_B$.]
		{\includegraphics[width=1.0\textwidth]{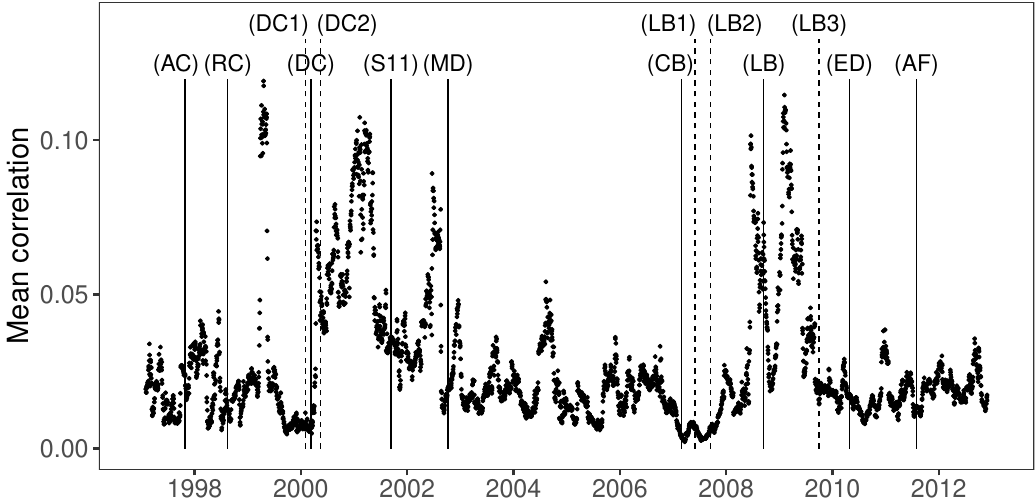}
		}\par
		\subfloat[\label{subfig:CorrRedRankMean}Mean correlation $\mean{\text{corr}}_L$.]
		{\includegraphics[width=1.0\textwidth]{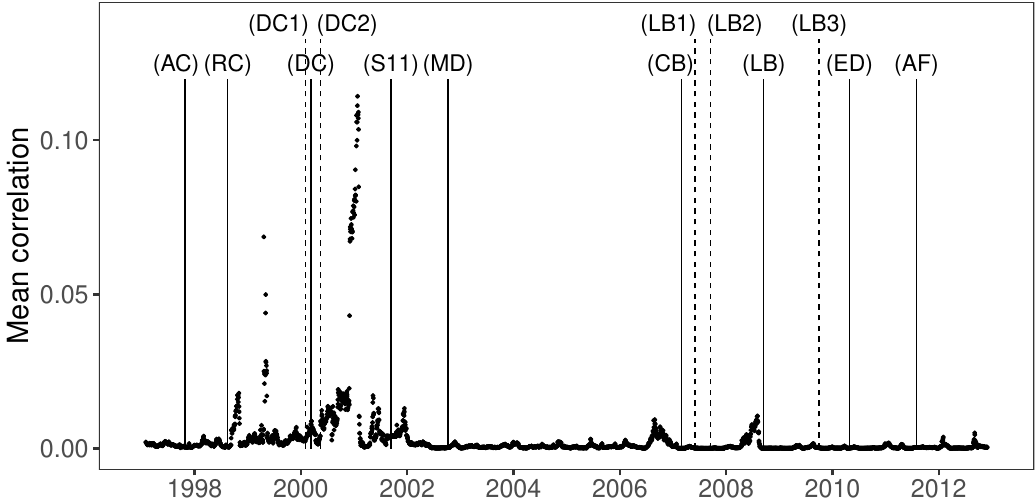}
		}
	\end{minipage}
	\caption{\label{subfig:Main:MeanValCov}Comparison of different mean values: \protect\subref{subfig:CovStandMean} mean covariance $\mean{\text{cov}}$, \protect\subref{subfig:CovSubstrDyadMean} mean covariance $\mean{\text{cov}}_B$,
	\protect\subref{subfig:CorrSubstrDyadMean} mean covariance $\mean{\text{cov}}_L$,
	 \protect\subref{subfig:CorrStandMean} mean correlation $\mean{\text{corr}}$,  
	 \protect\subref{subfig:CovRedRankMean} mean correlation $\mean{\text{corr}}_B$ and \protect\subref{subfig:CorrRedRankMean} mean correlation $\mean{\text{corr}}_L$.
		The beginning of the dot-com bubble burst is highlighted by historical event (DC) and the Lehman Brothers crash is highlighted by (LB). 	
		Further label explanations for historical events (lower row)  and estimated events (upper row) can be found in Tabs.~\ref{tab:FinancialCrises} and \ref{tab:EventsMarktStateTrans}. Each dot represents an interval of 42 trading days (\href{https://www.quandl.com/}{Data 
			from QuoteMedia via Quandl}).}
\end{figure*}

In Ref.~\cite{Heckens_2020}, we introduced the covariance approach and the correlation approach.
The covariance approach uses the standard covariance matrix, the correlation approach employs the standard correlation matrix,
where the term ``standard" refers to the original covariance and correlation matrix as obtained from the measured time series. 
In Sec.~\ref{sec:CovAppr}, we give a short introduction to the covariance approach, as well as in Sec.~\ref{sec:CorrAppr} to the correlation approach.

\subsection{\label{sec:CovAppr}Covariance approach}

The starting point is the evaluation of the standard covariance matrix
\begin{equation} \label{eqn:CovarianceMatAADagger}
	\Sigma = \frac{1}{T} A A^{\dagger} = \sum_{i=1}^K \kappa_i u_i u^{\dagger}_i \,.
\end{equation}
The $K \times T$ data matrix $A$ contains mean-normalized time series as rows.
Additionally, we apply a spectral decomposition to the standard covariance matrix.
Eigenvalues are denoted by $\kappa_i$ and eigenvectors are denoted by $u_i$.
The largest eigenvalue and the corresponding eigenvector can be interpreted as market part of $\Sigma$, while the other larger eigenvalues correspond to industrial sectors.
The smaller eigenvalues form a so-called bulk which contains to a large extent the noise in the time series. Their spectral density can be described by random matrix distributions~\cite{Marchenko_1967,Laloux_1999, Noh_2000, Gopikrishnan_2001, Plerou_2002, Song_2011, MacMahon_2015,  Stepanov_2015, Benzaquen_2017, potters2020first}.
Now, we subtract the dyadic matrix corresponding to the  largest eigenvalue
\begin{equation} \label{eqn:CovarianceMatSigmaB}
	\Sigma_B = \Sigma - \kappa_K u_K u^{\dagger}_K \,
\end{equation}
arriving at the matrix $\Sigma_B$ as an intermediate quantity.
We mention in passing that $\Sigma_B$ is a well-defined covariance matrix~\cite{Heckens_2020}.
Using the standard deviations ordered in the diagonal matrix
\begin{equation} \label{eqn:VolatilityMatrixB}
	\sigma_B =
	\textrm{diag} \left( \sigma_{B1} , \ldots , \sigma_{BK} 
	\right) \,,
\end{equation}
we define the reduced-rank correlation matrix in the covariance approach
\begin{equation} \label{eqn:CorrelationMatB}
	C_B = \left( \sigma_B \right)^{-1} \, \Sigma_B \, \left( \sigma_B \right)^{-1} \,.
\end{equation}
In Fig.~\subref*{subfig:CorrMatEntTimePeriodCovAppr}, the correlation matrix in the covariance approach is depicted for the entire 16 year period. It shows positively correlated block-diagonal entries corresponding to eleven industrial sectors and anti-correlations in the inter-sector structure.

\subsection{\label{sec:CorrAppr}Correlation approach}

Here, the starting point is the calculation of the standard correlation matrix
\begin{equation} \label{eqn:CorrelationMatMMDagger}
	C = \frac{1}{T} M M^{\dagger} = \sum_{i=1}^K \lambda_i x_i x^{\dagger}_i \,,
\end{equation}
where $M$ is a $K \times T$ data matrix whose rows are normalized to standard deviation one and mean value zero.
Eigenvalues are denoted by $\lambda_i$ and eigenvectors by $x_i$.
Analogously to Eq.~\eqref{eqn:CorrelationMatB}, by means of the matrix
\begin{equation} \label{eqn:CovarianceMatSigmaL}
	\Sigma_L = C - \lambda_K x_K x^{\dagger}_K \,
\end{equation}
and the diagonal matrix of standard deviations
\begin{equation} \label{eqn:VolatilityMatrixL}
	\sigma_L =
	\textrm{diag} \left( \sigma_{L1} , \ldots , \sigma_{LK} 
	\right) \,,
\end{equation}
we define the reduced-rank correlation matrix of the correlation approach 
\begin{equation} \label{eqn:CorrelationMatL}
	C_L = \left( \sigma_L \right)^{-1} \, \Sigma_L \, \left( \sigma_L \right)^{-1} \,.
\end{equation}
The reduced-rank correlation matrix in the correlation approach is depicted in Fig.~\subref*{subfig:CorrMatEntTimePeriodCorrAppr}. 
It looks very similar to the one in the
covariance approach for the entire 16 year period in Fig.~\subref*{subfig:CorrMatEntTimePeriodCovAppr}.
As shown in Ref.~\cite{Heckens_2020}, around the Lehman Brothers crisis, the reduced-rank correlation matrices for both approaches differ very much from each other for a one year time period.

\begin{figure}[!htb]
	\centering
	\includegraphics[width=1.0\columnwidth]{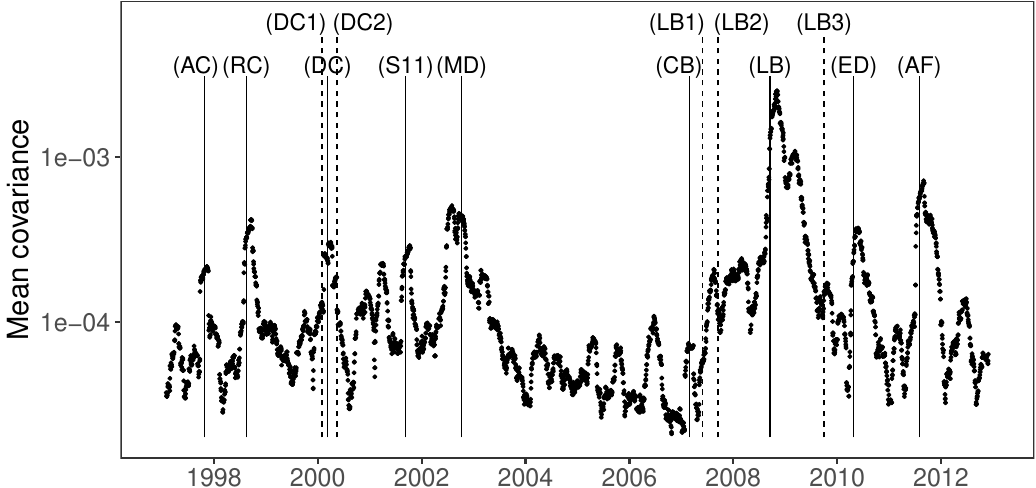}
	\caption{Mean covariance $\mean{\text{cov}}$ plotted on a logarithmic scale in order to visualize the increasing and decreasing mean covariances around crisis listed in Tabs~\ref{tab:FinancialCrises}. The other events can be found in Tab.~\ref{tab:EventsMarktStateTrans} (\href{https://www.quandl.com/}{Data from QuoteMedia 
			via Quandl}).}
	\label{fig:MeanCovStandLOG}
\end{figure}
\begin{figure*}[!htb]
	% \vspace{-1cm}
	\centering
		\subfloat[\label{subfig:DistMatCovAppr}]{
			\includegraphics[width=0.5\textwidth]{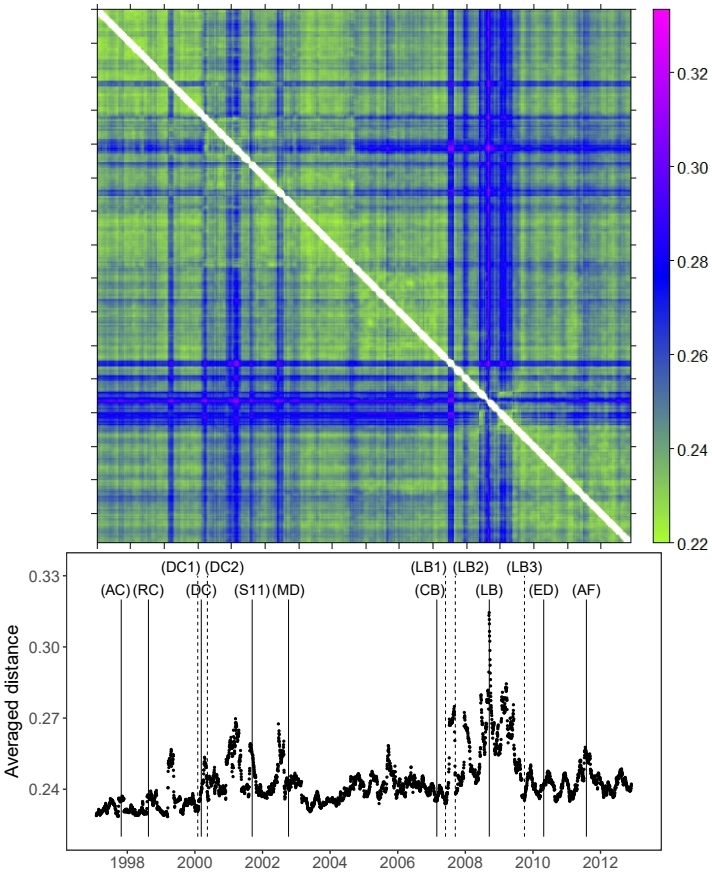}
		}%
		\subfloat[\label{subfig:DistMatCorrAppr}]{
			\includegraphics[width=0.5\textwidth]{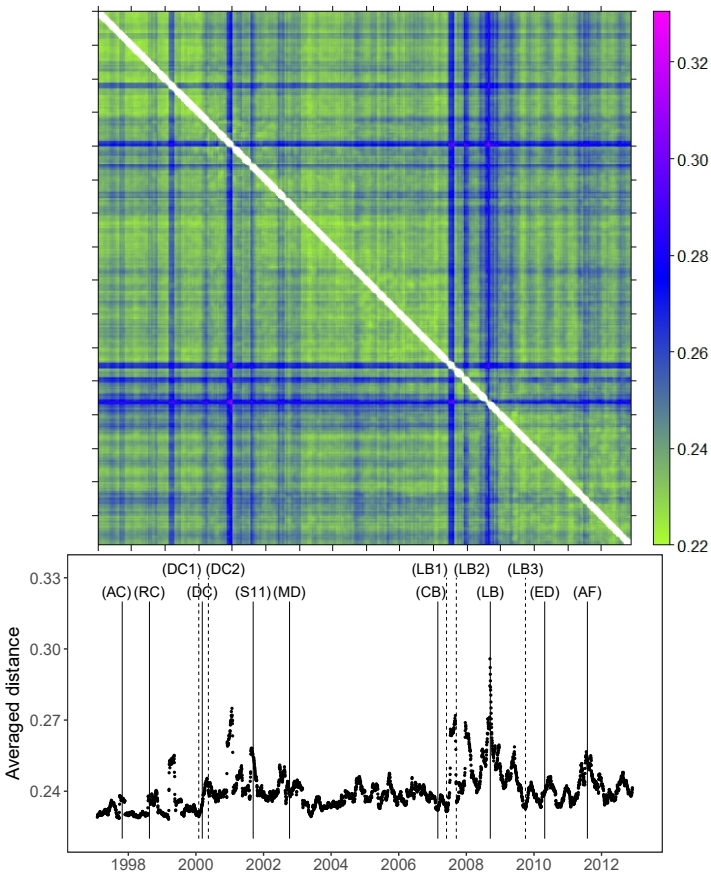}
		}%
	\caption{\label{subfig:Main:DistMatCovAppr} \protect\subref{subfig:DistMatCovAppr} Covariance approach: (top) distance matrix $\zeta^{\;\text{Eucl}}_B \left(t_a, t_b \right)$ and (bottom) averaged distance $\mean{\zeta}^{\;\text{Eucl}}_B \left(t_a\right)$;
	 \protect\subref{subfig:DistMatCorrAppr} Correlation approach: (top) 
	 distance matrix $\zeta^{\;\text{Eucl}}_L \left(t_a, t_b \right)$ and (bottom) averaged distance $\mean{\zeta}^{\;\text{Eucl}}_L \left(t_a\right)$.
	For the averaged distances, values smaller than 0.22 are excluded. Marks along margins of distance matrices indicate trading years.	
	Events in brackets are listed in Tabs.~\ref{tab:FinancialCrises} and \ref{tab:EventsMarktStateTrans} (\href{https://www.quandl.com/}{Data 
			from QuoteMedia via Quandl}).}
\end{figure*}

\section{\label{sec:DataAnalysisAndResults}Data Analysis and Results}

In Sec.~\ref{sec:ConceptOfMarketStates}, we briefly recapitulate our concept of market states.
In Sec.~\ref{sec:MeanCaluesDistanceMatricesAndAverageDistances}, we define mean values, distance matrices and averaged distances for the reduced-rank correlation matrices and visualize their temporal evolution.
We compare the time series of these mean values and averaged distances with historical and estimated events in Sec.~\ref{sec:HistoricalAndEstimatedEvents} in order to study
transitions of market states.
Other precursors are identified in Sec.~\ref{sec:OtherLongTermPrecursors}.
In Sec.~\ref{sec:MarketStateTransitionsAsLongTimePrecursorsOfEndogenousCrises}, we investigate
market states for the dot-com bubble burst and the pre-phase of the Lehman Brothers crises.
Importantly, we do not use post-crisis data.

\subsection{\label{sec:ConceptOfMarketStates}Concept of market states}

We identify quasi-stationary structures in the time-dependent correlation matrices. We refer to them as market states.
In previous works \cite{munnixIdentifyingStatesFinancial2012,Stepanov_2015,Rinn_2015,Pharasi_2018, Chetalova_2015,Chetalova_2015_2,
	papenbrockHandlingRiskonRiskoff2015,Qiu_2018,Heckens_2020,pharasi2020market,
	pharasi2020dynamics,Wang_2020,pharasi2021dynamics}, time periods of several decades are divided into so-called epochs (see~Sec.~\ref{sec:DataSet}).
Epochs are usually disjoint, \textit{i.e.} non-overlapping intervals. The fixed length of each epoch is typically one or two trading months.
For each epoch, we calculate a correlation matrix, a standard one or a reduced-rank one.
We group correlation matrices of the same kind by employing a clustering algorithm, the $k$-means clustering algorithm~\cite{Steinhaus1956:DivisionCorpMaterielEnParties,
	BallHall1965:ISODATA, MacQueen1967:SomeMethodsClassification,
	LLoyd1982:LeastSquaresQuantization, Jain_2010}.
We identify these groups as market states which comprise correlation matrices of a similar structure. 
Wo do that for the standard and for the reduced-rank correlation matrices which show striking differences.
In particular, we find longer life times in the latter ones.
The market operates in a state for a certain time, then jumps to another one and yet another one and also might return to a state. Put differently, the states emerge, exist for some time, reappear and eventually disappear.

\subsection{\label{sec:MeanCaluesDistanceMatricesAndAverageDistances}Mean values, distance matrices and averaged distances}

Since we want to scrutinize what happens at the transitions of the market states we use overlapping $T_{\text{ep}} = 42$~trading days to find signals in our data potentially being connected to such transitions.
Relevant signals may be found in the mean values of all reduced-rank correlation matrices elements and in the distance matrices derived from the reduced-rank correlation matrices.
Investigating time series with a moving interval of $T_{\text{ep}} = 42$~trading days involves noise in the corresponding observables such as volatilites, covariances or correlations. Numerous analysis especially in the past decade showed
that this is justified by revealing dynamics in financial markets which otherwise would stay hidden (see~Ref.~\cite{marti2021review}).
Kenett et al.~\cite{kenett2010dynamics} analyzed the US stock markets in a period from 2002 to 2009.
These authors found that 22 trading days are approximately the lowest bound for a 
trade-off between noise and a smoothed dynamics for so-called normalized correlation matrices.

We set up $K \times T_{\text{ep}}$ data matrices $A$ and $M$ where $K=250$ is the number of stocks and $T_{\text{ep}} = 42$ trading days is the length of one interval.
The matrices are computed  on a 42 trading day sliding window which is shifted forward by one trading day and moved over the 4026 trading days of the return time series. In total, we compute 3984 data matrices $A$ and 3984 data matrices $M$.
According to Sec.~\ref{sec:ConstructionOfReducedRankMatrices}, we calculate the standard covariance matrix $\Sigma$, the covariance matrix $\Sigma_B$, the reduced-rank correlation matrix $C_B$ in the covariance approach, the standard correlation matrix $C$, the covariance matrix $\Sigma_L$ in the covariance approach and the reduced-rank correlation matrix $C_L$ in the correlation approach.
The latter matrices are singular and their elements are strongly influenced by noise.

However, it is quite difficult to separate the dynamics and noise introduced on such short time windows.
To make statements about the system-relevant dynamics for matrices $\Sigma$, $\Sigma_B$ $C_B$, $C$, $\Sigma_B$ and $C_L$ with \mbox{$T_{\text{ep}} = 42$~trading days} we assume that the ten largest eigenvalues for each matrix contain almost all of the information on the industrial sectors. We take into account only the ten largest eigenvalues similar to a cut-off from the noise-dressed bulk for matrices with \mbox{$T \gg K$}~\cite{Marchenko_1967,Laloux_1999, Noh_2000, Gopikrishnan_2001, Plerou_2002, Song_2011, MacMahon_2015,  Stepanov_2015, Benzaquen_2017, potters2020first}.
We define the fraction
\begin{equation} \label{eqn:Fraction}
	f_{\Sigma n} = \frac{\sum_{i=1}^{n} \kappa_{K+1-i} }{\sum_{i=1}^{10} \kappa_{K+1-i}} \,,
\end{equation}
for the covariance matrix $\Sigma$ (cf.~Ref.~\cite{Wang_2011}). The fraction $f_{\Sigma n}$ describes how much variance of all 250 time series correspond to the $n$ largest eigenvalues related to the variance of the ten largest eigenvalues.

Figure~\ref{subfig:Main:Fractions} depicts the fractions $n=1,\ldots,4$ for all six matrices.
For $\Sigma$ and $C$ in Figs.~\ref{subfig:CovStandFraction} and \ref{subfig:CorrStandFraction}, the four largest eigenvalues contain at least 53\,\% and on average 74\,\% of the variance of the ten largest eigenvalues. 
After 2007/2008, the dynamics of the selected stocks is dominated by the largest eigenvalues of $\Sigma$ and $C$.
Subtracting the dyadic matrix corresponding to the largest eigenvalue reveals a completely different dynamics in Figs.~\ref{subfig:CovSubstrFraction}, \ref{subfig:CorrSubstrFraction}, \ref{subfig:CovRedRankFraction} and \ref{subfig:CorrRedRankFraction}.
The four largest eigenvalues contain at least 46\,\% and on average 55\,\% of the variance of the ten largest eigenvalues.
The dynamics is not dominated by the largest eigenvalues of $\Sigma_B$, $\Sigma_L$, $C_B$ and $C_L$ after 2007/2008.
The time evolution of the curves show a certain stability in time indicating a stable correlation structure and corroborating our results for the market state analysis from Ref.~\cite{Heckens_2020}. 
Remarkably, the four curves in Figs.~\ref{subfig:CovSubstrFraction}, \ref{subfig:CorrSubstrFraction}, \ref{subfig:CovRedRankFraction} and \ref{subfig:CorrRedRankFraction} feature a similar time evolution for the entire time period reflecting the coherent behavior of the industrial sectors.

As a next step, the mean value of all matrix elements for every matrix is computed.
For the covariance matrix $\Sigma$, we take the average of all covariance matrix elements
\begin{equation} \label{eqn:MeanCovariances}
	\mean{\text{cov}} = \frac{1}{K^2} \sum_{i,j = 1}^K \Sigma_{ij} \,.
\end{equation}
Correspondingly, we define the mean covariance $\mean{\text{cov}}_B$ in the covariance approach and the mean covariance $\mean{\text{cov}}_L$ in the correlation approach. Since we include the diagonal elements in the average in Eq.~\eqref{eqn:MeanCovariances}, we analogously
define the average for the standard correlation matrix $C$ as
\begin{equation} \label{eqn:MeanCorrelations}
    \mean{\text{corr}} = \frac{1}{K^2} \sum_{i,j = 1}^K C_{ij} \,.
\end{equation}
For the reduced-rank correlation matrix $C_B$ and $C_L$, the mean correlation is denoted by $\mean{\text{corr}}_B$ and $\mean{\text{corr}}_L$, respectively.
Finally, we generate time series of the mean covariances $\mean{\text{cov}}$, $\mean{\text{cov}}_B$, $\mean{\text{cov}}_L$ and the mean correlations $\mean{\text{corr}}$, $\mean{\text{corr}}_B$, $\mean{\text{corr}}_L$.
Each of these time series comprises 3984 mean values.

In Figs.~\subref*{subfig:CovStandMean}, \subref*{subfig:CovSubstrDyadMean} and~\subref*{subfig:CovRedRankMean},
the time evolutions of the mean values $\mean{\text{cov}}$,
$\mean{\text{cov}}_B$ and $\mean{\text{corr}}_B$ belonging to the covariance approach are displayed on a linear scale.
This is compared to the time evolutions of the mean values $\mean{\text{corr}}$, $\mean{\text{cov}}_L$ and $\mean{\text{corr}}_L$ in Figs.~\subref*{subfig:CorrStandMean}, \subref*{subfig:CorrSubstrDyadMean} and~\subref*{subfig:CorrRedRankMean}
corresponding to the correlation approach on a linear scale as well.
Each data matrix, correlation matrix and mean value receive a time
stamp which corresponds to the center of a 42 trading day interval and
which is represented by a black dot
in Fig.~\ref{subfig:Main:MeanValCov}.
A logarithmic scale in Fig.~\ref{fig:MeanCovStandLOG} for mean covariance $\mean{\text{cov}}$ facilitates a comparison with its peaks.
Both mean covariances $\mean{\text{cov}}$ and $\mean{\text{cov}}_B$ are positive for the entire 16 year time period.
In contrast to the mean correlation $\mean{\text{corr}}$, the mean covariance $\mean{\text{cov}}$ shows no trend to larger mean values.
Larger mean values in the covariance and correlation approach correspond to reduced-rank correlation matrices in which a sector such as the financial sector or a subgroup of stocks show sometimes strong anti-correlation to the other industrial sectors (cf.~Ref.~\cite{Heckens_2020}). For smaller mean covariances $\mean{\text{cov}}$ and correlations $\mean{\text{corr}}$, we usually find larger peaks in $\mean{\text{corr}}_B$ and $\mean{\text{corr}}_L$, respectively.
Although the mean correlations $\mean{\text{corr}}_B$ and $\mean{\text{corr}}_L$ can be small in contrast to $\mean{\text{corr}}$, 
this does not imply that these are artifacts due to noise as
we effectively take the average of $K(K-1)/2 = 31125$ correlation matrix elements, taking the symmetry and the unities on the diagonals of correlation matrices into account. For the mean covariances, the variances (squared volatilities) are on the diagonals and are time-dependent quantities.

We now turn to measuring distances between correlation matrices.	
As we use 
$k$-means clustering in our market state analysis which employs the Euclidean metric~\cite{Steinhaus1956:DivisionCorpMaterielEnParties,
	BallHall1965:ISODATA, MacQueen1967:SomeMethodsClassification,
	LLoyd1982:LeastSquaresQuantization, Jain_2010}, we introduce a distance matrix -- as defined for the standard correlation matrix in Refs.~\cite{munnixIdentifyingStatesFinancial2012, Stepanov_2015} -- by calculating the pairwise distance between two reduced-rank correlation matrices in the covariance approach
\begin{align} \label{eqn:ScaledEuclidDist}
	\zeta^{\;\text{Eucl}}_B \left(t_a, t_b \right) &= \frac{\sqrt{ \sum_{i,j}
			\left( C_{Bij}( t_a ) - C_{Bij}( t_b )    
			\right)^2 }} {K}
	\,.
\end{align}
Correspondingly, in the correlation approach, we define the
Euclidean distance $\zeta^{\;\text{Eucl}}_L \left(t_a, t_b \right)$.
Rows and columns are labeled by the
indices $t_a = 1, \ldots, 3984$ and $t_b = 1, \ldots, 3984$.
Thus, we obtain distance matrices of dimension $3984 \times 3984$. 
Due to the normalization with $K$ in Eq.~\eqref{eqn:ScaledEuclidDist} we may compare
distance matrices
for another selection of stocks.

In Figs.~\subref*{subfig:DistMatCovAppr} and~\subref*{subfig:DistMatCorrAppr} the distance matrices calculated according to Eq.~\eqref{eqn:ScaledEuclidDist} are displayed.
The larger the distance, the more dissimilar are two reduced-rank correlation matrices. 
The distance matrices derived from the standard correlation matrices show their highest values for
financial crashes \cite{munnixIdentifyingStatesFinancial2012}.
We observe a completely different dynamics for the distance matrices  $\zeta^{\;\text{Eucl}}_B \left(t_a, t_b \right)$ and $\zeta^{\;\text{Eucl}}_L \left(t_a, t_b \right)$.
Quasi-stationary periods of distances with values of approximately 0.24 are followed by less stable periods of larger distances with values larger than 0.26.
The distance matrix $\zeta^{\;\text{Eucl}}_B \left(t_a, t_b \right)$ for the covariance approach shows a much more pronounced structure than that for the correlation approach $\zeta^{\;\text{Eucl}}_L \left(t_a, t_b \right)$.

We introduce a new time series -- the \textit{averaged distance} -- to analyze the market state transitions in more detail. Due to the characteristic stripes in Figs.~\subref*{subfig:DistMatCovAppr} and~\subref*{subfig:DistMatCorrAppr}, it is beneficial to take the average of the rows of a distance matrix since these stripes might contain information on possible market state transitions.
Using the distance matrix $\zeta^{\;\text{Eucl}}_B \left(t_a, t_b \right)$ in~Eq.~\eqref{eqn:ScaledEuclidDist},
we calculate the averaged distance as
\begin{align} \label{eqn:AveragedScaledEuclidDist}
	\mean{\zeta}^{\;\text{Eucl}}_B(t_a) = \frac{1}{t_c} \sum_{t^{\prime}_b=1}^{t_c} \zeta^{\;\text{Eucl}}_B \left(t_a, t^{\prime}_b \right)
	\,.
\end{align}
keeping the row index $t_a$ fixed and summing up to a specific column index $t_c$.
The column index $t_b^{\prime} =1$ corresponds to the trading date 1997-01-31
and $t_c = 484$
to the trading date \mbox{1998-12-31}. 
Thus, we obtain a side profile for each distance matrix mapping important structural information into a one-dimensional plot.
The averaged distance facilitates the comparison with historical events. 
We do not analyze $\mean{\zeta}^{\;\text{Eucl}}_B(t_a)$ in dependence of $t_c$.
Analogously, we define the averaged distance $\mean{\zeta}^{\;\text{Eucl}}_L(t_a)$ for the correlation approach.
The averaged distances
are depicted in Figs.~\subref*{subfig:DistMatCovAppr} for the covariance approach and in Fig.~\subref*{subfig:DistMatCorrAppr} for the correlation approach.

In Figs.~\subref*{subfig:DistMatCovAppr} and~\subref*{subfig:DistMatCorrAppr}, we cut off distance matrix elements with values smaller than 0.22. Thereby, we exclude distances between overlapping reduced-rank correlation matrices in Eq.~\eqref{eqn:AveragedScaledEuclidDist}. 
By including values smaller than 0.22 we would calculate systematically lower values for the averaged distance in Eq.~\eqref{eqn:AveragedScaledEuclidDist} from $t_b^{\prime} = 1$ to $t_c$
compared to the time period from $t_c$ to $t_d = 3984$.
We would create an artificial jump at $t_c$.

\subsection{\label{sec:HistoricalAndEstimatedEvents}Historical and estimated events}

We wish to compare the time evolution of the mean values and averaged distances with events which are on the one hand historical crisis events listed in~Tab.~\ref{tab:FinancialCrises} and on the other hand events in~Tab.~\ref{tab:EventsMarktStateTrans} estimated from the averaged distances in Sec.~\ref{sec:MeanCaluesDistanceMatricesAndAverageDistances}.
Those estimated events are connected to market state transitions 
and are highlighted as dashed lines in Figs.~\ref{subfig:Main:MeanValCov},~\ref{fig:MeanCovStandLOG} and~\ref{subfig:Main:DistMatCovAppr}.

The mean covariance $\mean{\text{cov}}$ in Fig.~\subref*{subfig:CovStandMean} shows its highest peaks for the Lehman Brothers crash (LB) whereas the mean correlation $\mean{\text{corr}}$ in Fig.~\subref*{subfig:CorrStandMean} has its largest value for the August 2011 stock market fall (AF).
We want to emphasize that the trading day when the NASDAQ Composite stock market index peaked is labeled by (DC) \cite{wiki:2021:DOTCOMBUBBLE}. Shortly after this event, the dot-com bubble bursted.

Our new analysis corroborates our results in Ref.~\cite{Heckens_2020}: 
Using reduced-rank correlation matrices, in covariance or correlation approach, 
exogenous effects are likely to be efficiently separated from endogenous ones. 
This affects all stocks and appears in $\mean{\text{cov}}$ and  $\mean{\text{corr}}$
as crisis events (cf.~Tab.~\ref{tab:FinancialCrises}).
The mean correlations $\mean{\text{corr}}_B$ and $\mean{\text{corr}}_L$ show a non-stationary behavior with sometimes even high values, especially in the case of the covariance approach (cf.~Figs.~\subref*{subfig:CovRedRankMean} and \subref*{subfig:CorrRedRankMean}).
We additionally calculated the mean covariances $\mean{\text{cov}}_B$ and $\mean{\text{cov}}_L$
which both show a separation from the historical events in their temporal behavior as well (cf.~Figs.~\subref*{subfig:CovSubstrDyadMean} and~\subref*{subfig:CorrSubstrDyadMean}).

The averaged distances in Figs.~\subref*{subfig:DistMatCovAppr} and~\subref*{subfig:DistMatCorrAppr}
facilitate the identification of signals in the distance matrix which are connected to market state transitions.
The market state transitions occur for the dot-com bubble burst roughly between (DC1) and (DC2) and for the pre-phase of the Lehman Brothers crash between (LB1) and (LB2) (cf.~Tab.~\ref{tab:EventsMarktStateTrans}). The duration of the market state transition period 
in the vicinity of historical event (DC) is 73~trading days and prior to (LB) 74 trading days.
The market state transition for the Lehman Brothers crisis market state appears in mid-2007 as it was observed in Ref.~\cite{zheng2012changes} for the absolute changes of the largest eigenvalues of the standard correlation matrices. The presumable cause was the freezing of the Interbank market \cite{Investopedia}.
This is a precursor signal exclusively for (LB).

The dashed lines (LB1) and (LB3) around (LB) show a high agreement with the recession period of December~2007~-~June~2009~\cite{wiki:2021:LISTUSRECESSIONS}.
Information on such recession periods is provided by the National Bureau of Economic Research (NBER) for the US
economy. Apparently, we are able to detect connections to this recession period using reduced-rank correlation matrices on US stock markets.
It is possible that the observables in our study are influenced by the recession period from March 2001 - November 2001 as well.

Additionally, Figs.~\subref*{subfig:DistMatCovAppr} and~\subref*{subfig:DistMatCorrAppr}, also make visible the historical events (AC), (RC), (S11) and (AF) and potentially (ED) in the averaged distances.
The difference to the dot-com bubble burst and the Lehman Brothers crisis is that these four to five crises are not accompanied by
larger outburst in the mean covariances $\mean{\text{cov}}_B$, $\mean{\text{cov}}_L$ and the mean correlations $\mean{\text{corr}}_B$ and $\mean{\text{corr}}_B$
in their direct vicinity (see~Fig.~\ref{subfig:Main:MeanValCov}).
In spite of removing to a large extent
exogenous contributions in the mean covariances  $\mean{\text{cov}}_B$ and $\mean{\text{cov}}_L$ and mean correlations $\mean{\text{corr}}_B$ and $\mean{\text{corr}}_L$ we find that the averaged distances
still contain exogenous information of several financial crises.

\begin{table}[!htb]
	\centering
	\caption{\label{tab:FinancialCrises}
		Historical events taken 
		from~\cite{wiki:2021:LISTSTOCKSCRASHES}.
	}
	\begin{tabular}{cl@{\hspace{1em}}c}
		\toprule
		Label &
		Crisis &
		\multicolumn{1}{c}{Date}
		\\
		&
		&
		\multicolumn{1}{c}{(Year-Month-Day)} \\
		\midrule
		(AC) & Asian financial crisis & 1997-10-27 \\
		(RC) & Russian financial crisis & 1998-08-17 \\
		(DC) & Dot-com bubble (before burst) & 2000-03-10 \\
		(S11) & September 11th & 2001-09-11 \\
		(MD) & Stock market downturn of 2002 & 2002-10-09 \\
		(CB) & Chinese stock bubble & 2007-02-27 \\
		(LB) & Lehman Brothers crash & 2008-09-16 \\
		(ED) & European debt crisis & 2010-04-27 \\
		(AF) & August 2011 stock markets fall & 2011-08-01 \\
		\bottomrule
	\end{tabular}
\end{table}
\begin{table}[!htb]
	\centering
	\caption{\label{tab:EventsMarktStateTrans}
		Dates for events estimated from the averaged distances introduced in Sec.~\ref{sec:MeanCaluesDistanceMatricesAndAverageDistances}.
		First to fourth label highlight dates for begin and end of market state transitions (\href{https://www.quandl.com/}{Data 
			from QuoteMedia via Quandl}).
	}
	\begin{tabular}{l@{\hspace{0.3em}}l@{\hspace{0.2em}}c}
		\toprule
		Label &
		Description &
		\multicolumn{1}{c}{Date}
		\\
		&
		&
		\multicolumn{1}{c}{(Year-Month-Day)} \\
		\midrule
		(DC1) & Start of market state transition  & 2000-02-01 \\
		& for dot-com bubble burst   &  \\
		(DC2) & End of market state transition  & 2000-05-15 \\
		& for dot-com bubble burst  &            \\
		(LB1) & Start of market state transition  & 2007-06-01 \\
		& for Lehman Brothers crisis  &     \\
		(LB2) & End of market state transition & 2007-09-15 \\
		& for Lehman Brothers crisis &  \\
		(LB3) & End of & 2009-10-01 \\
		& Lehman Brothers crisis period &  \\
		\bottomrule
	\end{tabular}
\end{table}

\subsection{\label{sec:OtherLongTermPrecursors}Other long-term precursors}

The periods of the dot-com bubble burst between (DC1) and (DC2) and the pre-phase of the Lehman Brothers crash between (LB1) and (LB2) start with very low values of $\mean{\text{corr}}_B$ accompanied by an increasing mean covariance $\mean{\text{cov}}$ for (LB), also depicted on a logarithmic scale in Fig.~\ref{fig:MeanCovStandLOG}.
These very low values of $\mean{\text{corr}}_B$ coincide with market state transitions.
In the case of the dot-com bubble burst a sudden mean correlation outburst in $\mean{\text{corr}}_B$ appears at (DC1).
For the Lehman pre-phase, however, the outburst of $\mean{\text{corr}}_B$ does not happen at (LB1) but shortly before (LB).
The low values for the mean correlation $\mean{\text{corr}}_B$ are precursor signals for (DC) and (LB).

The sudden decreases in the mean covariances $\mean{\text{cov}}_B$ and $\mean{\text{cov}}_L$ and in the mean correlations $\mean{\text{corr}}_B$ and $\mean{\text{corr}}_L$ at (LB) indicate a connection
to the Lehman Brothers crash whereas the
mean covariance $\mean{\text{cov}}$ as a measure for the entire market risk and the mean correlation $\mean{\text{corr}}$ have peaks. The financial sector crisis spread over to the entire market, leading to a market-wide crash. 
This can be interpreted as a spill-over effect.
Therefore, $\mean{\text{cov}}_B$, $\mean{\text{cov}}_L$, $\mean{\text{corr}}_B$ and $\mean{\text{corr}}_L$ show features of potential measures for systemic risk.
We do not observe such a spill-over effect for the dot-com bubble.
The largest peaks in the averaged distances in Figs.~\subref*{subfig:DistMatCovAppr} and \subref*{subfig:DistMatCorrAppr} at (LB) coincide with lower mean correlations $\mean{\text{corr}}_B$ and $\mean{\text{corr}}_L$. 
This spill-over effect is a precursor signal exclusively for (LB).

Comparing the averaged distances in Figs.~\subref*{subfig:DistMatCovAppr} and~\subref*{subfig:DistMatCorrAppr} and the time evolution of the mean correlations $\mean{\text{corr}}_B$ and $\mean{\text{corr}}_L$ in Figs.~\subref*{subfig:CovRedRankMean} and~\subref*{subfig:CorrRedRankMean} we see that
changes in the averaged distances occur prior to the changes in the mean correlation at a market state transition to the crisis period for (LB).
In the pre-phase of the Lehman Brothers crisis, the mid-2007 peak discussed in Sec.~\ref{sec:HistoricalAndEstimatedEvents} does not coincide with a larger change in mean correlations. The spill-over effect appears later.
In the case of the dot-com bubble burst, 
we observe for the correlation approach between (DC1) and (DC2) that a change in the averaged distance is accompanied by an outburst of the mean correlation $\mean{\text{corr}}_B$.

\subsection{\label{sec:MarketStateTransitionsAsLongTimePrecursorsOfEndogenousCrises}Market state transitions as long-term precursors of financial crises}

\begin{figure*}[!htb]
	% \vspace{-1cm}
	\centering
	\begin{minipage}{0.5\textwidth}
		\subfloat[\label{subfig:CovApprLB_1}7 epochs before (LB).]{
			\includegraphics[width=0.85\textwidth]{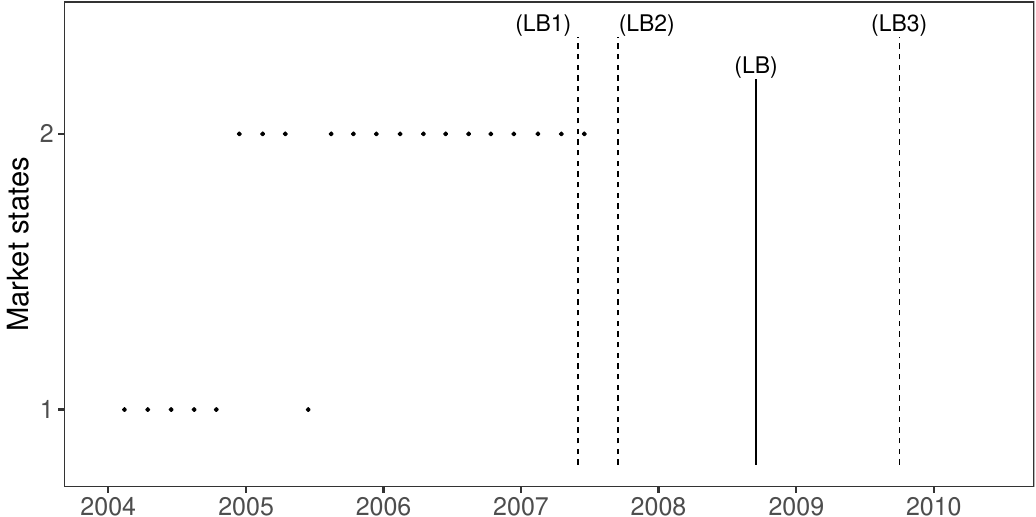}
		}\par
		\subfloat[\label{subfig:CovApprLB_2}6 epochs before (LB).]{
			\includegraphics[width=0.85\textwidth]{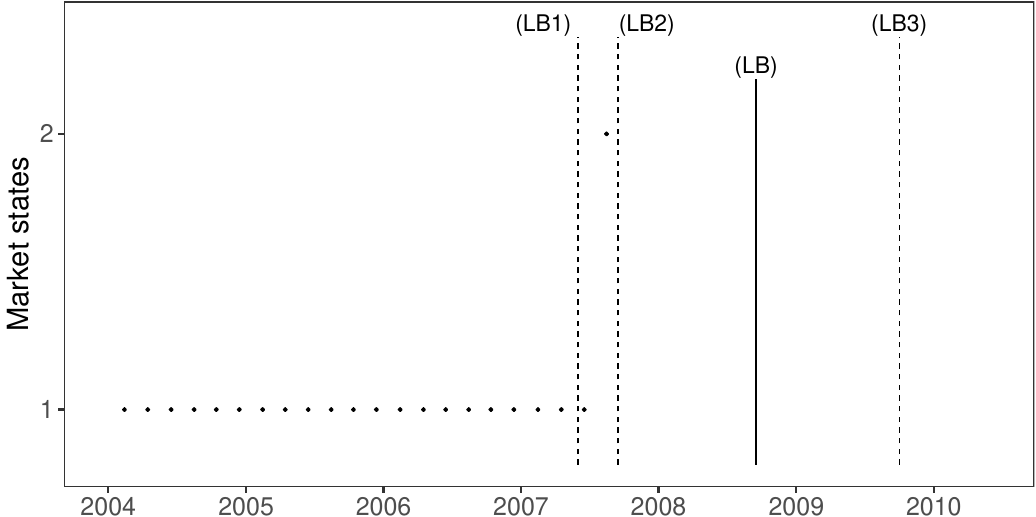}
		}\par
		\subfloat[\label{subfig:CovApprLB_3}3 epochs before (LB).]{
			\includegraphics[width=0.85\textwidth]{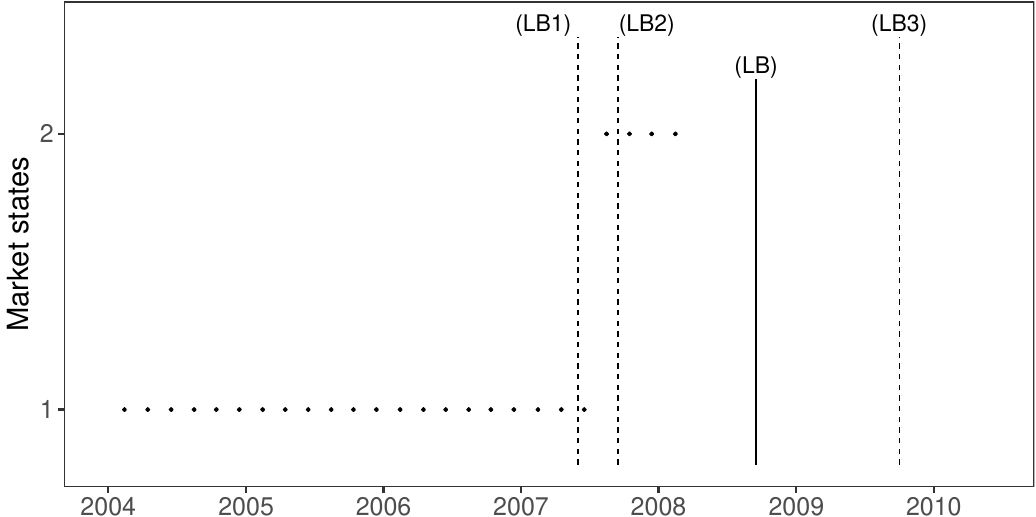}
		}\par
		\subfloat[\label{subfig:CovApprLB_4}2 epochs before (LB).]{
			\includegraphics[width=0.85\textwidth]{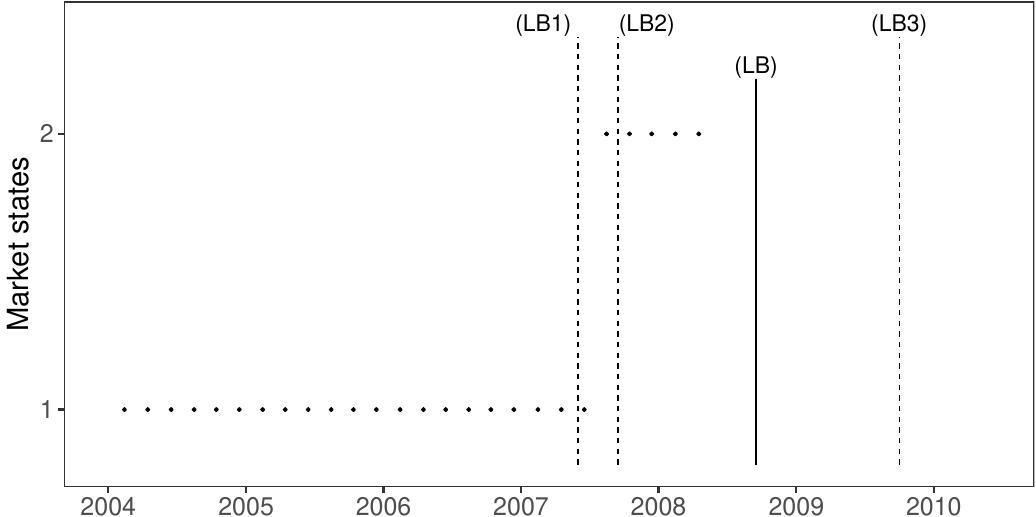}
		}\par
		\subfloat[\label{subfig:CovApprLB_5}1 epochs before (LB).]{
			\includegraphics[width=0.85\textwidth]{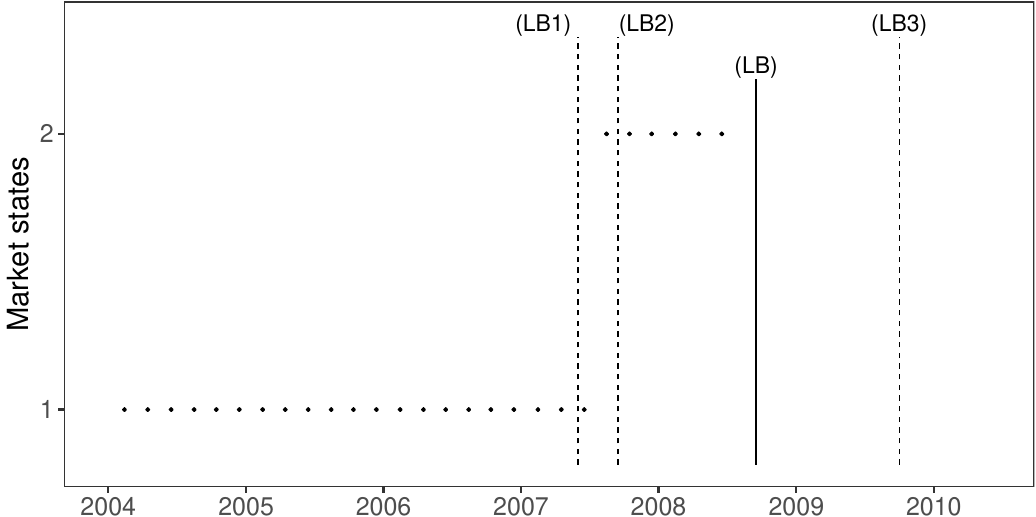}
		}
	\end{minipage}%
	\begin{minipage}{0.5\textwidth}
		\subfloat[\label{subfig:CorrApprLB_1}7 epochs before (LB).]{
			\includegraphics[width=0.85\textwidth]{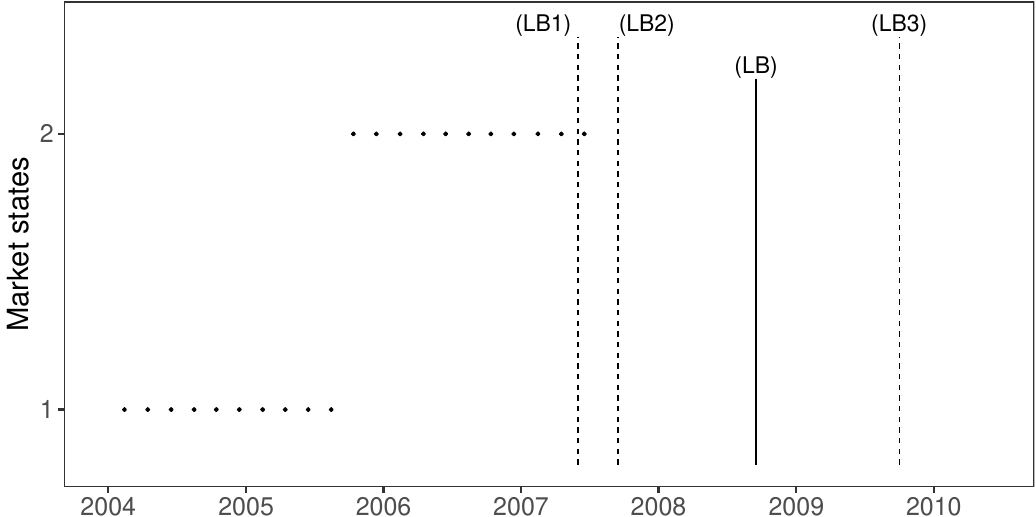}
		}\par
		\subfloat[\label{subfig:CorrApprLB_2}6 epochs before (LB).]{
			\includegraphics[width=0.85\textwidth]{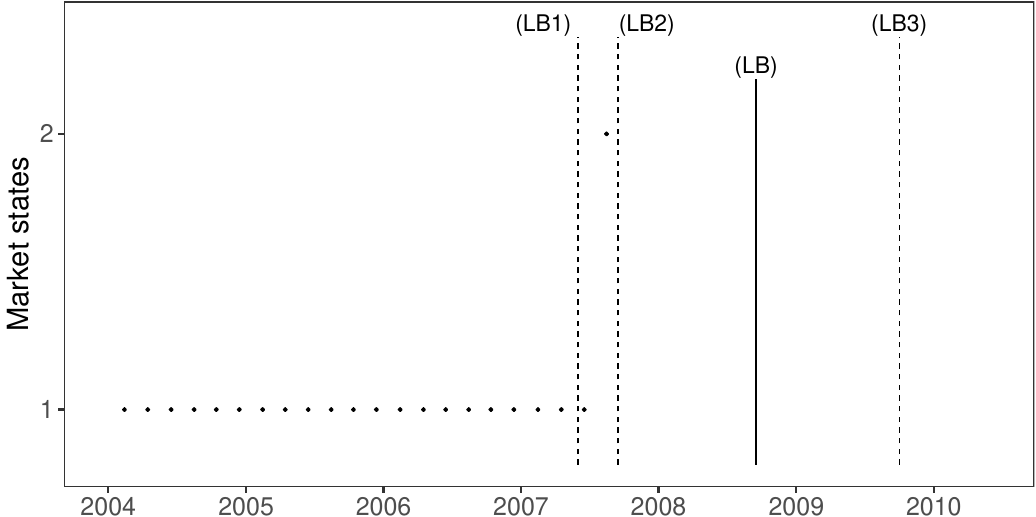}
		}\par
		\subfloat[\label{subfig:CorrApprLB_3}3 epochs before (LB).]{
			\includegraphics[width=0.85\textwidth]{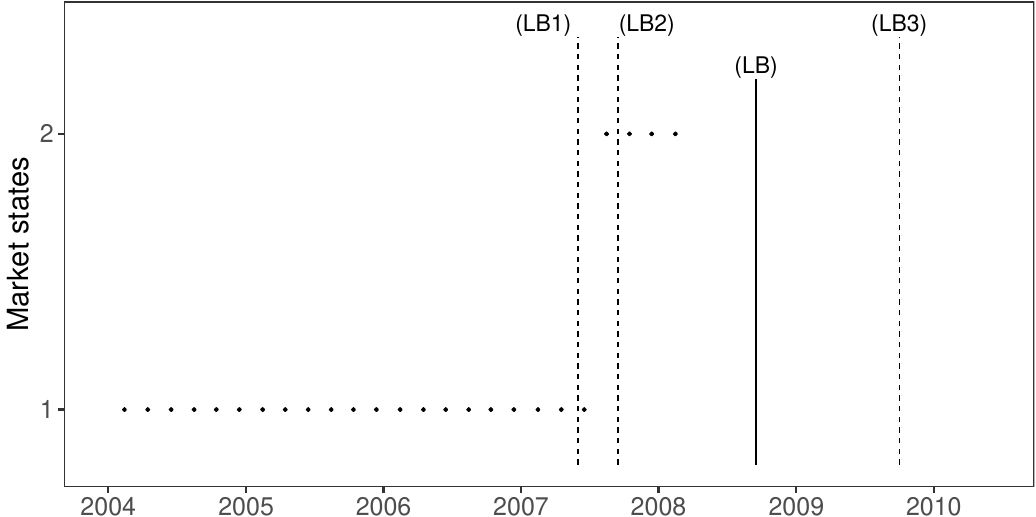}
		}\par
		\subfloat[\label{subfig:CorrApprLB_4}2 epochs before (LB).]{
			\includegraphics[width=0.85\textwidth]{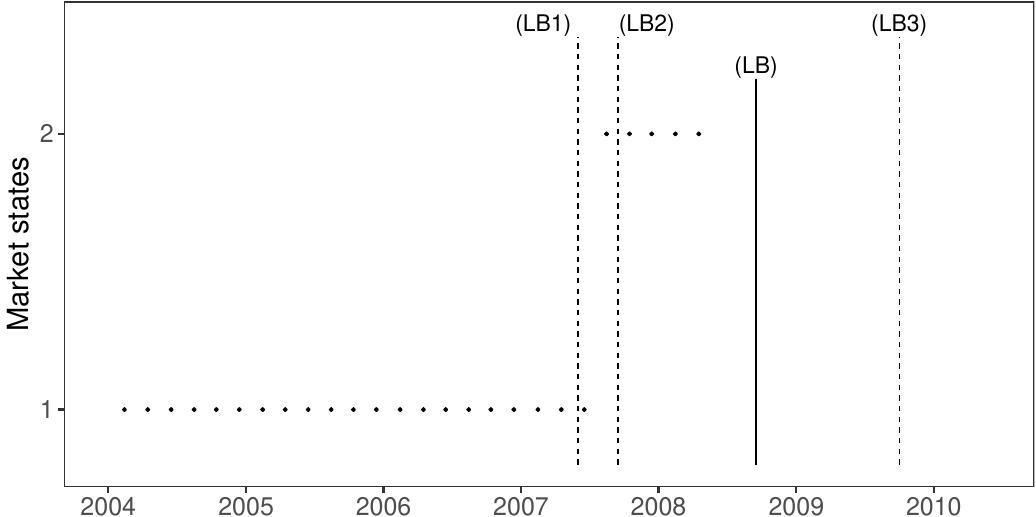}
		}
		\par
		\subfloat[\label{subfig:CorrApprLB_5}1 epochs before (LB).]{
			\includegraphics[width=0.85\textwidth]{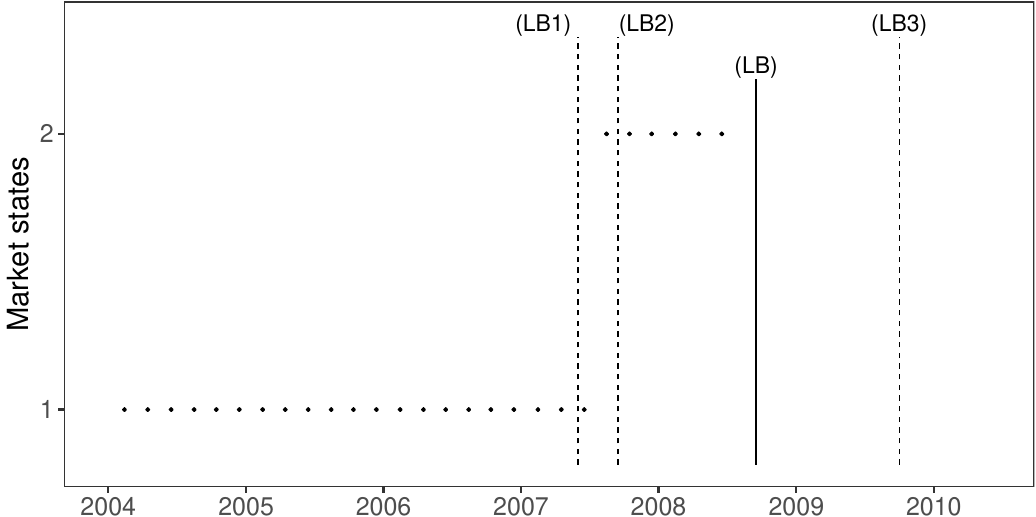}
		}
	\end{minipage}
	\caption{\label{subfig:Main:CovApprLB}Snapshots of the market state analysis for the reduced-rank correlation matrices in the \protect\subref{subfig:CovApprLB_1}-\protect\subref{subfig:CovApprLB_5} covariance approach and \protect\subref{subfig:CorrApprLB_1}-\protect\subref{subfig:CorrApprLB_5} correlation approach compared to estimated events around the Lehman Brothers crash (LB)  (for lower row, see~Tab.~\ref{tab:FinancialCrises}).
		For all snapshots, the first epoch is fixed.
		The clustered period starts for all snapshots at 2004-01-15.	
		The estimated events in the upper row (LB1), (LB2) and (LB3) can be found in Tab.~\ref{tab:EventsMarktStateTrans}.	
		Every dot stands for an epoch of 42 trading days. From subplots \protect\subref{subfig:CovApprLB_1} to \protect\subref{subfig:CorrApprLB_5}, the number of the remaining epochs before (LB) is specified   (\href{https://www.quandl.com/}{Data 
			from QuoteMedia via Quandl}).}
\end{figure*}
\begin{figure*}[!htb]
	% \vspace{-1cm}
	\centering
	\begin{minipage}{1.0\textwidth}
		\subfloat[\label{subfig:CovApprLB_1_Corr}7 epochs before (LB).]{
			\includegraphics[width=0.3\textwidth]{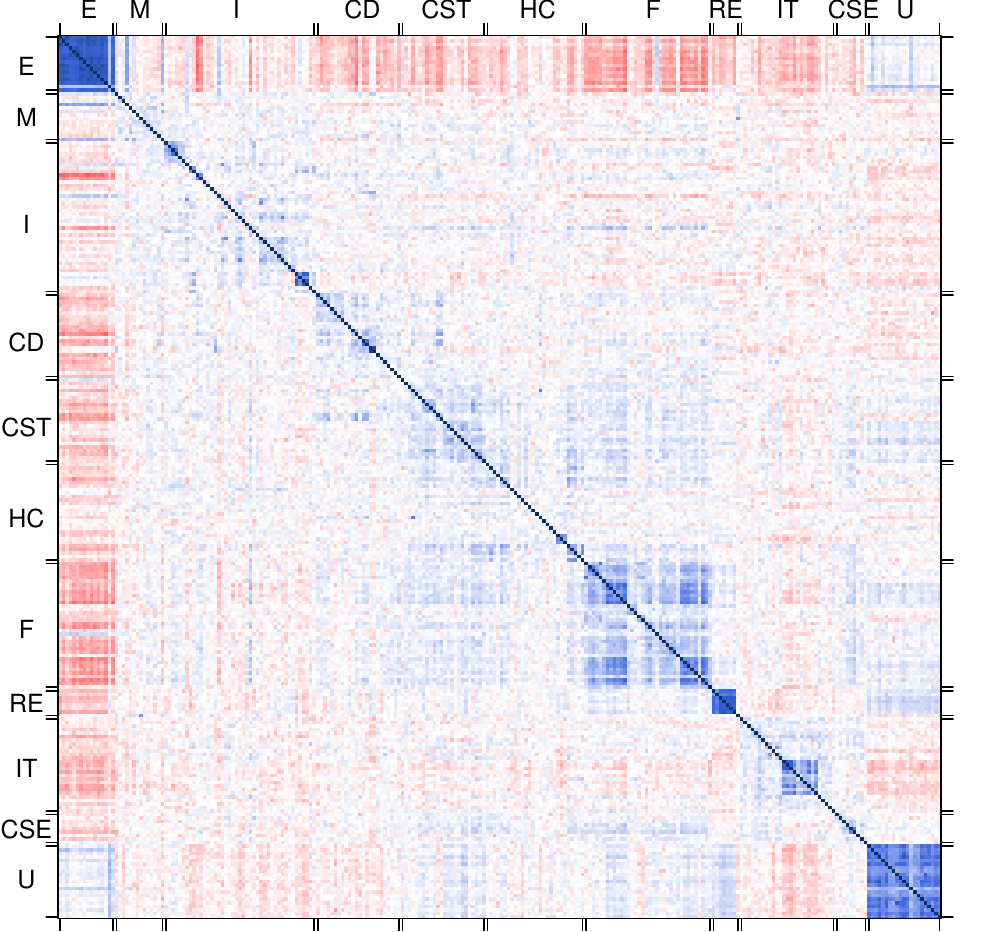}
		}
		\subfloat[\label{subfig:CovApprLB_2_Corr}6 epochs before (LB).]{
			\includegraphics[width=0.3\textwidth]{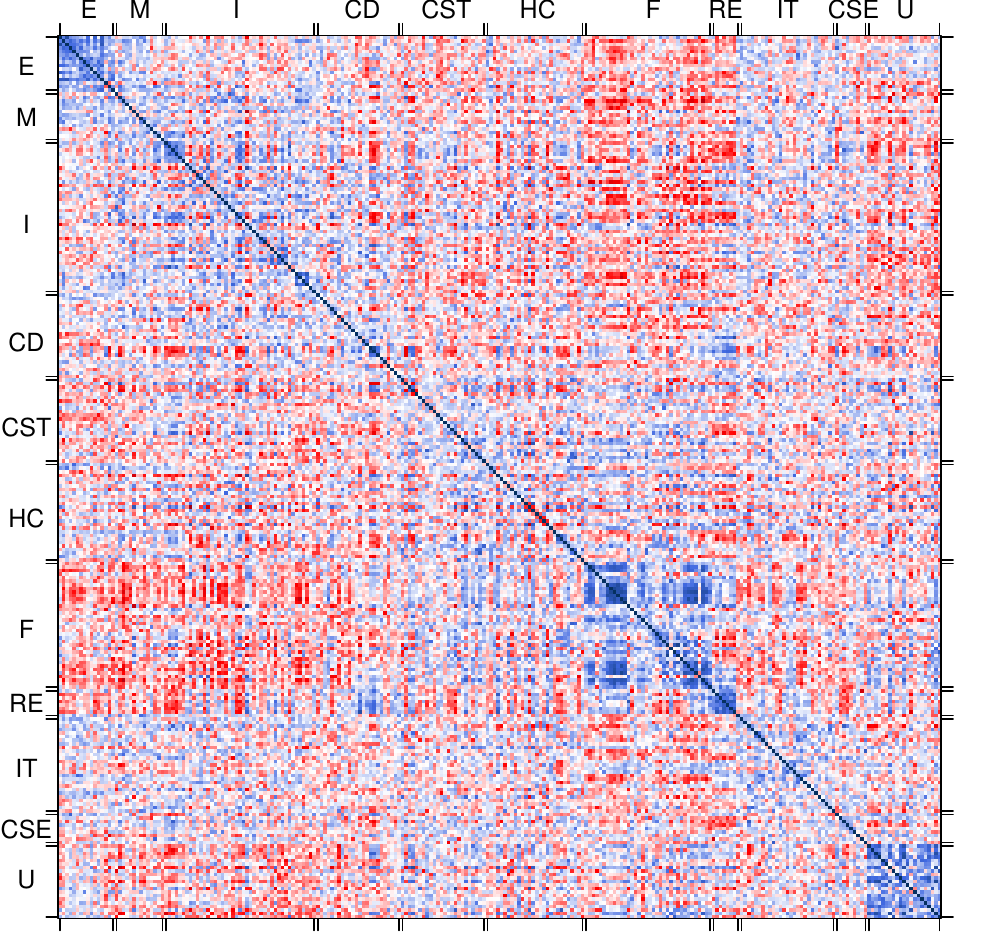}
		}
		\subfloat[\label{subfig:CovApprLB_3_Corr}3 epochs before (LB).]{
			\includegraphics[width=0.3\textwidth]{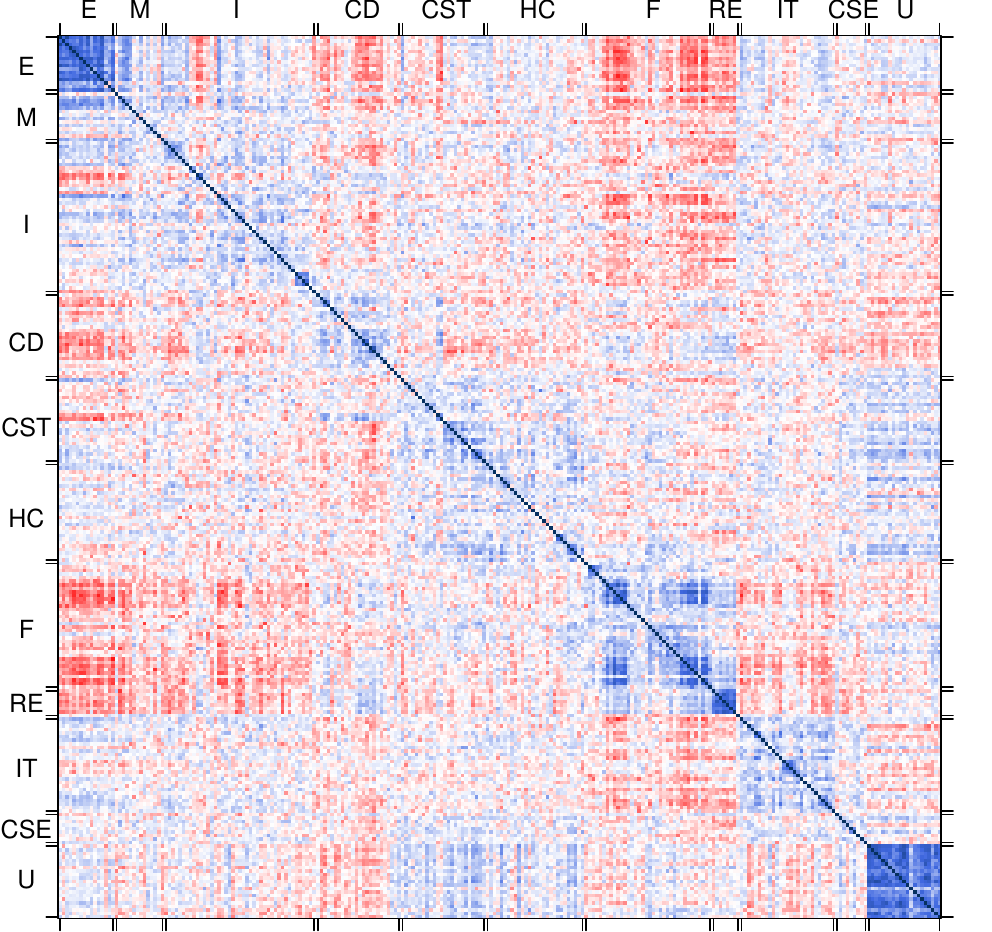}
		}
	\end{minipage}%
	\\
	\begin{minipage}{1.0\textwidth}
		\subfloat[\label{subfig:CorrApprLB_1_Corr}7 epochs before (LB).]{
			\includegraphics[width=0.3\textwidth]{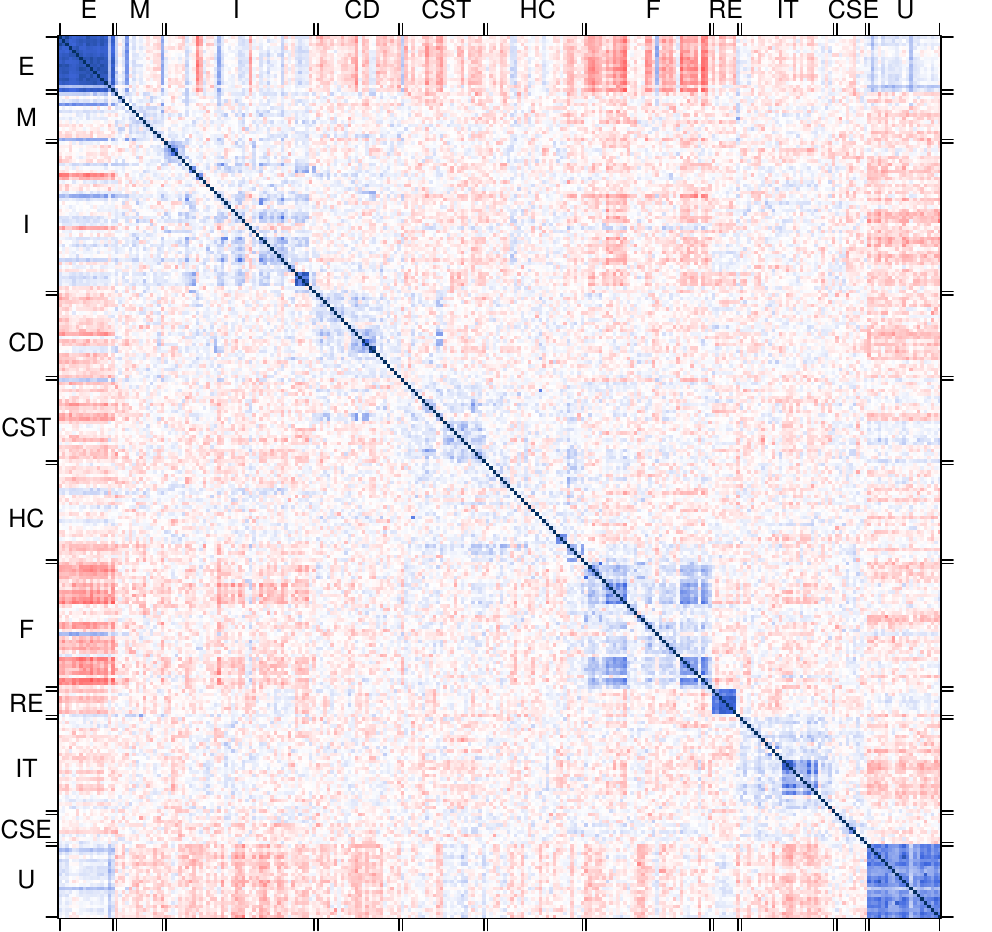}
		}
		\subfloat[\label{subfig:CorrApprLB_2_Corr}6 epochs before (LB).]{
			\includegraphics[width=0.3\textwidth]{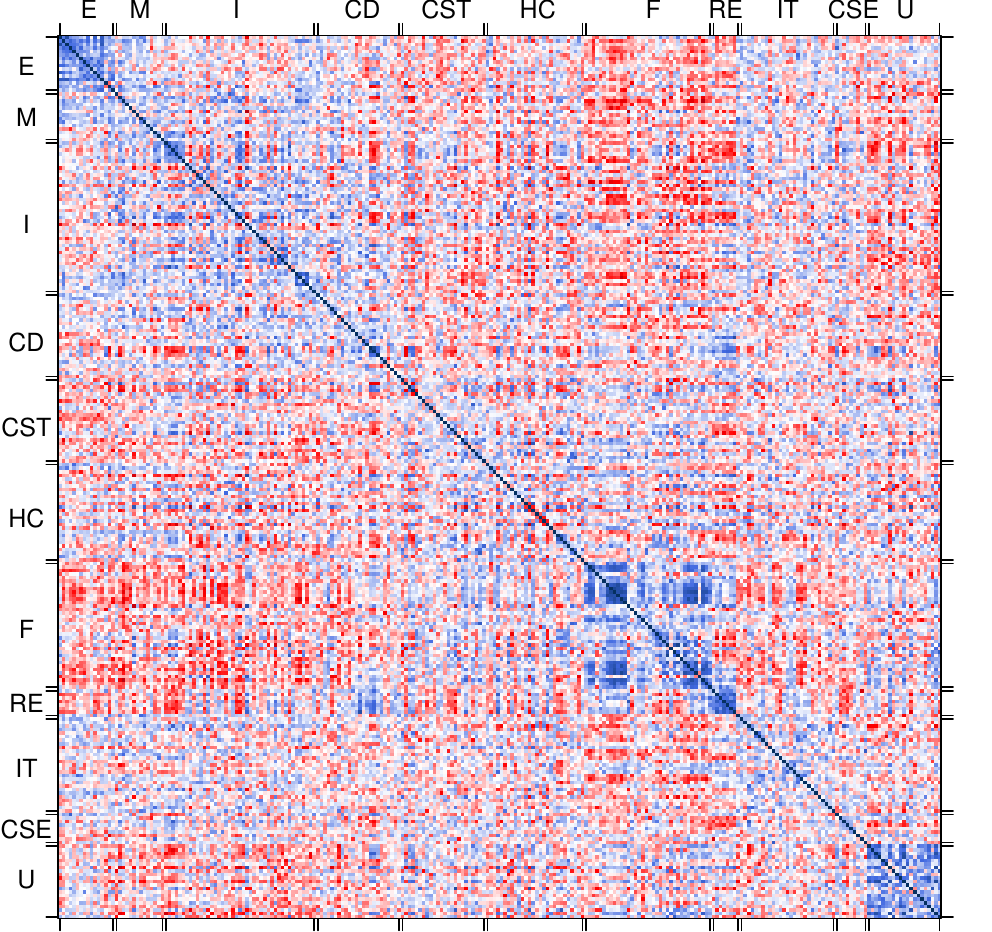}
		}
		\subfloat[\label{subfig:CorrApprLB_3_Corr}3 epochs before (LB).]{
			\includegraphics[width=0.3\textwidth]{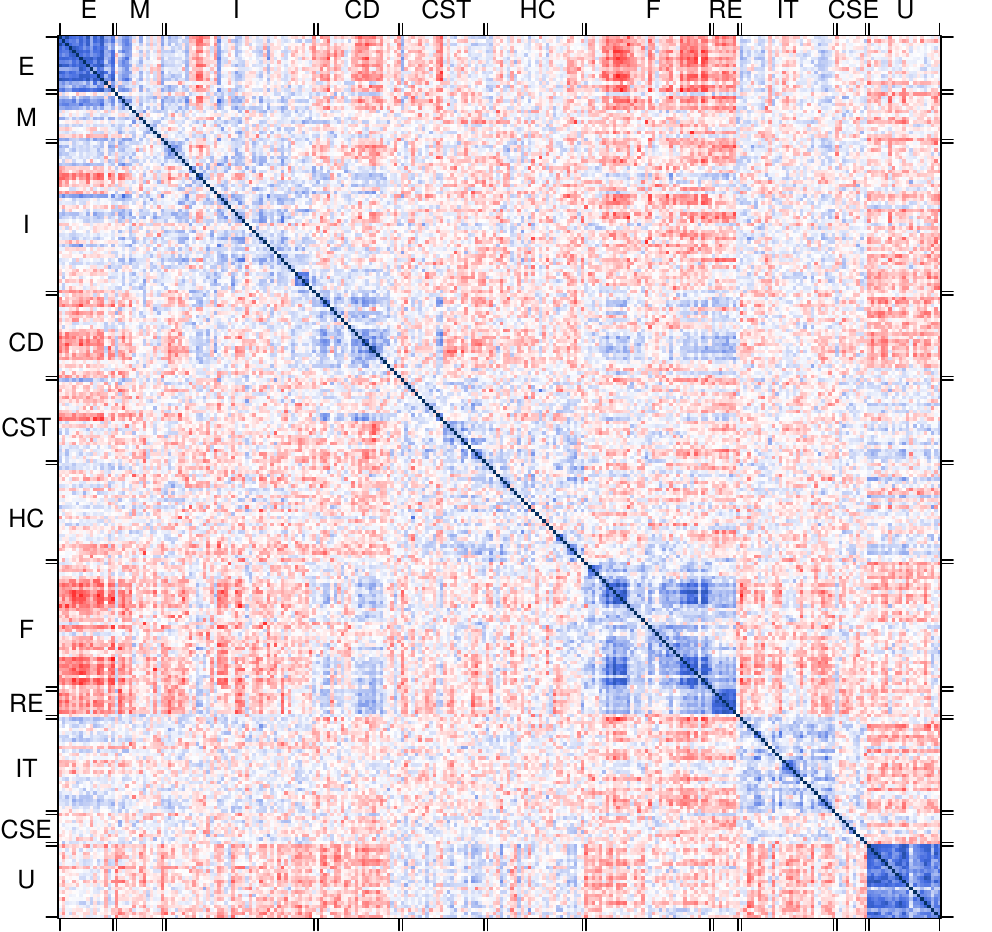}
		}
	\end{minipage}
	\caption{\label{subfig:Main:CorrApprLB_Corr}Typical second market states ($K = 250$ stocks) corresponding to the snapshots before the Lehman Brothers crash marked by historical event (LB) (cf.~Fig.~\ref{subfig:Main:CovApprLB}) for the reduced-rank correlation matrices in the \protect\subref{subfig:CovApprLB_1_Corr}-\protect\subref{subfig:CovApprLB_3_Corr} covariance approach and \protect\subref{subfig:CorrApprLB_1_Corr}-\protect\subref{subfig:CorrApprLB_3_Corr} correlation approach.
		From subplots \protect\subref{subfig:CovApprLB_1_Corr} to \protect\subref{subfig:CorrApprLB_3_Corr}, the number of the remaining epochs before (LB) is specified.
		A typical market state is obtained by taking the element-wise average of the correlation matrices of the respective market state. Capital Letters indicate industrial sectors (see~Tab.~\ref{tab:GICS})
		(\href{https://www.quandl.com/}{Data 
			from QuoteMedia via Quandl}).}
\end{figure*}

\begin{figure*}[!htb]
	% \vspace{-1cm}
	\centering
	\begin{minipage}{0.5\textwidth}
		\subfloat[\label{subfig:CovApprDC_1}0 epochs after (DC).]{
			\includegraphics[width=0.85\textwidth]{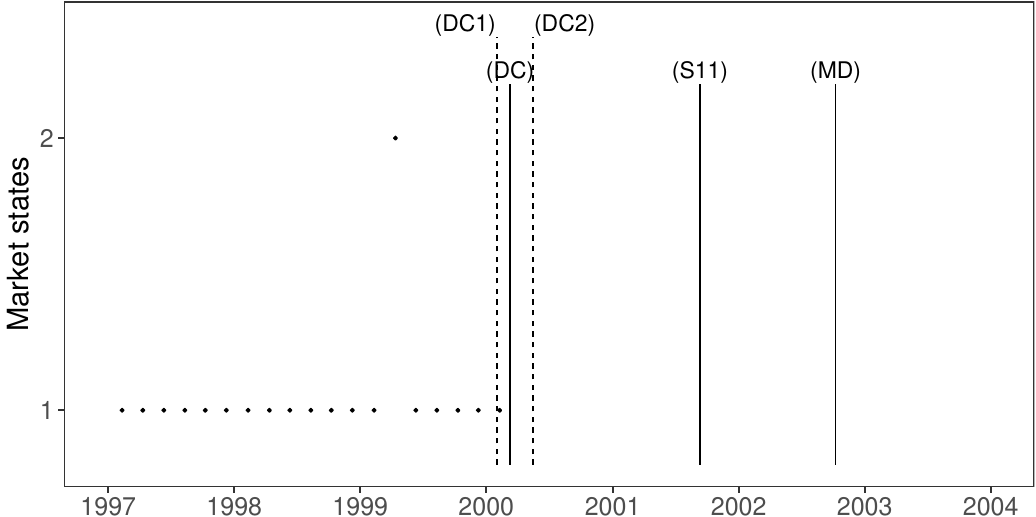}
		}\par
		\subfloat[\label{subfig:CovApprDC_2}2 epochs after (DC).]{
			\includegraphics[width=0.85\textwidth]{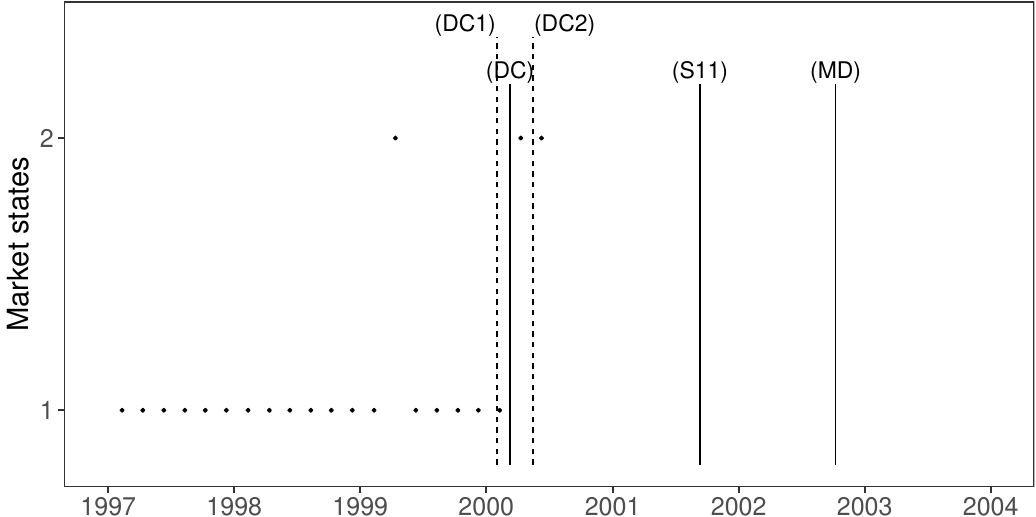}
		}\par
		\subfloat[\label{subfig:CovApprDC_3} 3 epochs after (DC).]{
			\includegraphics[width=0.85\textwidth]{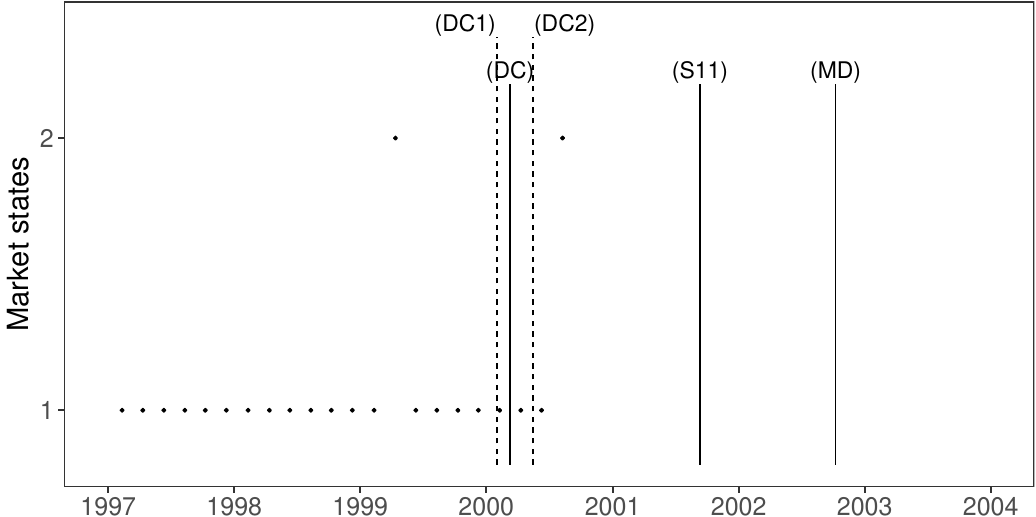}
		}\par
		\subfloat[\label{subfig:CovApprDC_4}4 epochs after (DC).]{
			\includegraphics[width=0.85\textwidth]{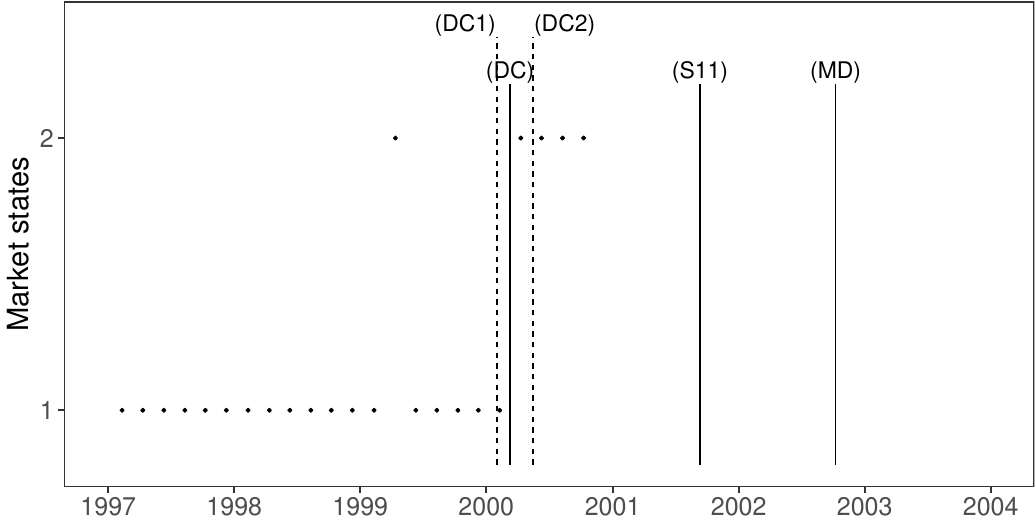}
		}\par
		\subfloat[\label{subfig:CovApprDC_5} 8 epochs after (DC).]{
			\includegraphics[width=0.85\textwidth]{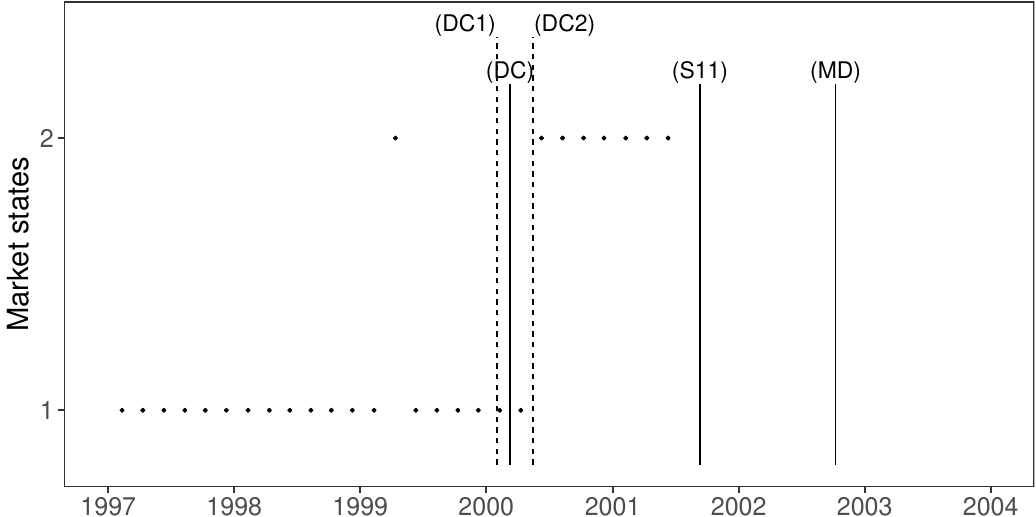}
		}
	\end{minipage}%
	\begin{minipage}{0.5\textwidth}
		\subfloat[\label{subfig:CorrApprDC_1}0 epochs after (DC).]{
			\includegraphics[width=0.85\textwidth]{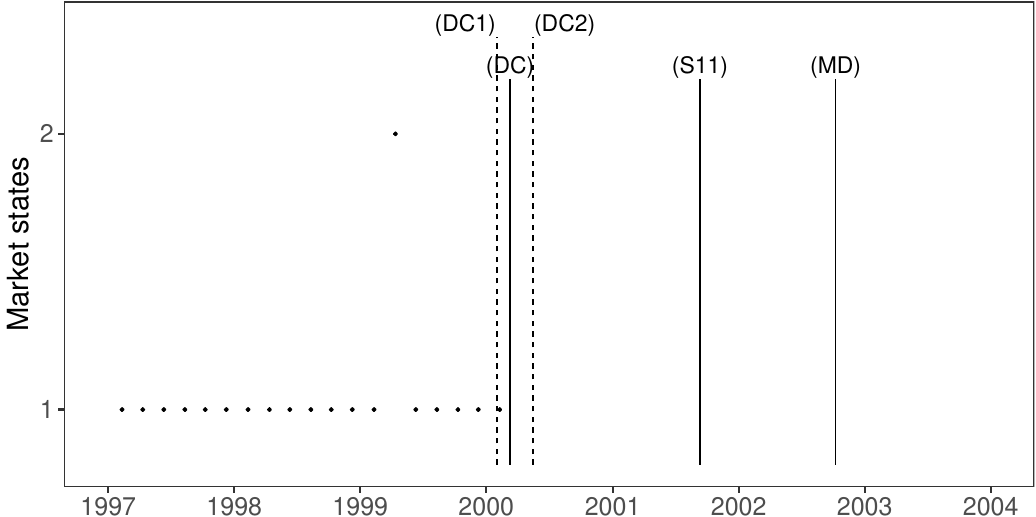}
		}\par
		\subfloat[\label{subfig:CorrApprDC_2}6 epochs after (DC).]{
			\includegraphics[width=0.85\textwidth]{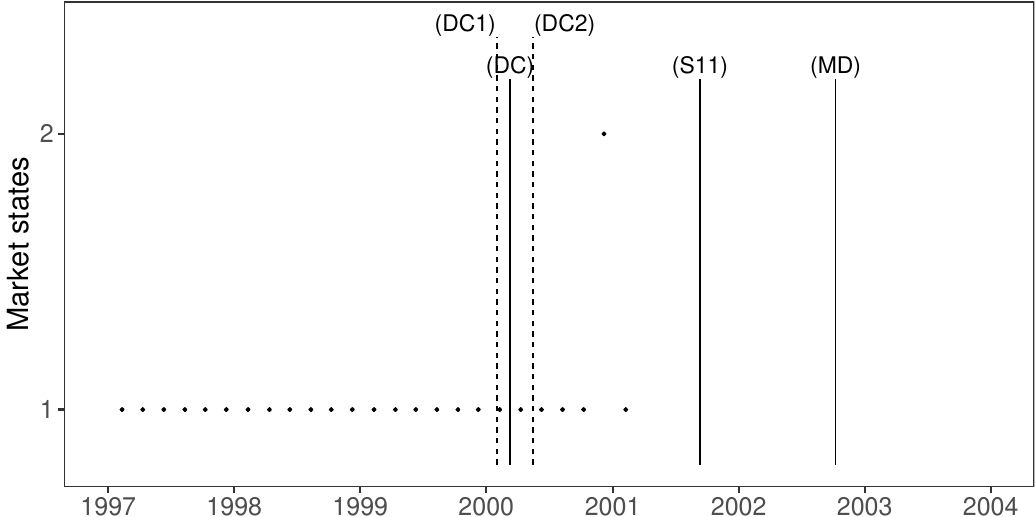}
		}\par
		\subfloat[\label{subfig:CorrApprDC_3}7 epochs after (DC).]{
			\includegraphics[width=0.85\textwidth]{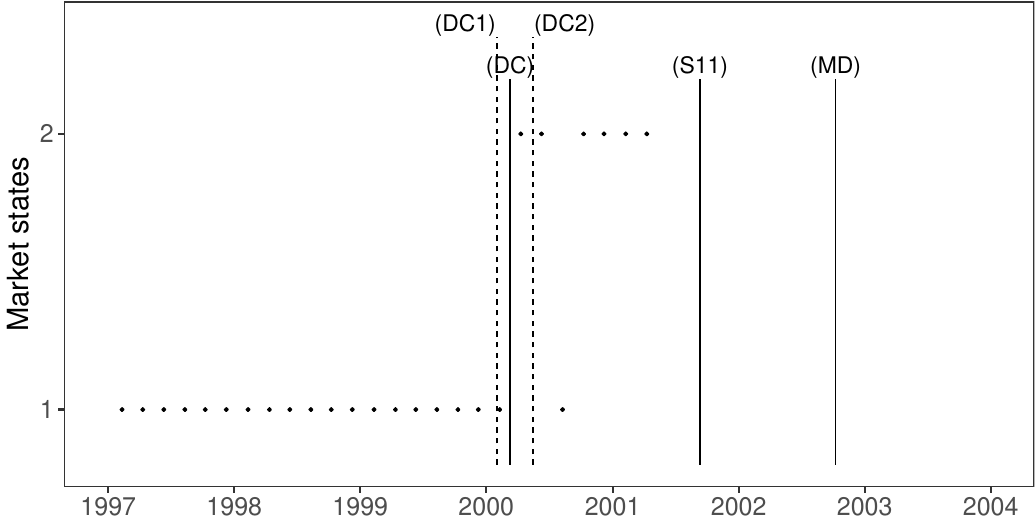}
		}
	\end{minipage}
	\caption{\label{subfig:Main:CovApprDC}Snapshots of the market state analysis for the reduced-rank correlation matrices in the \protect\subref{subfig:CovApprDC_1}-\protect\subref{subfig:CovApprDC_5} covariance approach and \protect\subref{subfig:CorrApprDC_1}-\protect\subref{subfig:CorrApprDC_3} correlation approach compared to estimated events at the beginning of the dot-com bubble burst marked by historical event (DC)  (for lower row, see~Tab.~\ref{tab:FinancialCrises}).
		For all snapshots, the first epoch is fixed.
		The clustered period starts for all snapshots at 1997-01-13.	
		The estimated events in the upper row (DC1) and (DC2) can be found in Tab.~\ref{tab:EventsMarktStateTrans}.	
		Every dot stands for an epoch of 42 trading days.
		From subplots \protect\subref{subfig:CovApprDC_1} to \protect\subref{subfig:CorrApprDC_3}, the number of the epochs after (DC) is specified   (\href{https://www.quandl.com/}{Data 
			from QuoteMedia via Quandl}).}
\end{figure*}

\begin{figure}[!htb]
	\centering
	\includegraphics[width=1.0\columnwidth]{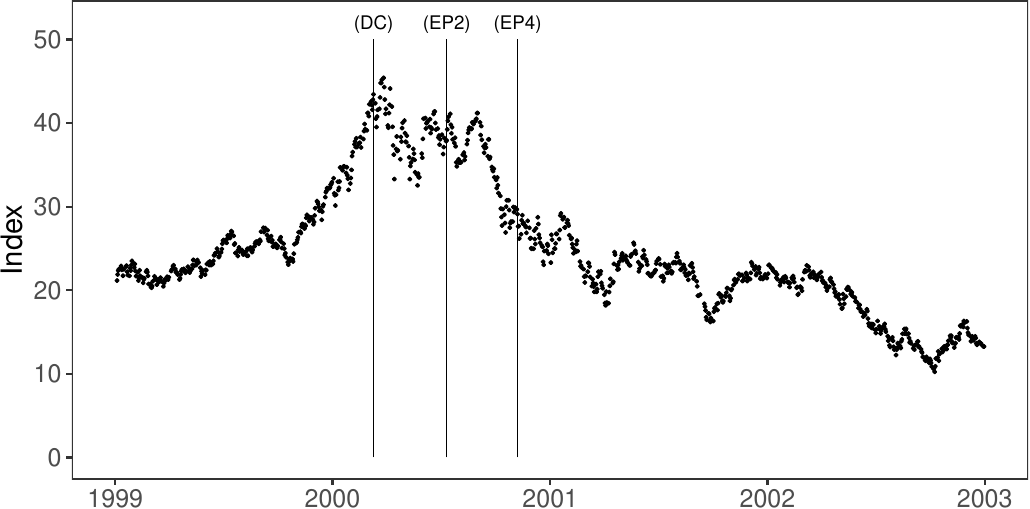}
	\caption{Self-constructed index as average of the adjusted daily closing prices (see~Sec.~\ref{sec:DataSet}) of the 27 IT stocks (see~Tab.~\ref{tab:GICS} and Tab.~\ref{tab:OverviewSP500} in App.~\ref{sec:ListStocks}).	
		The dates for the end of the epochs are for (EP2) 2000-07-11 (two epochs after (DC)) and for (EP4) 2000-11-07 (four epochs after (DC)) (\href{https://www.quandl.com/}{Data from QuoteMedia 
			via Quandl}).}
	\label{fig:IndexDC}
\end{figure}
\begin{figure*}[!htb]
	% \vspace{-1cm}
	\centering
	\begin{minipage}{1.0\textwidth}
		\subfloat[\label{subfig:CovApprDC_1_Corr}0 epochs after (DC).]{
			\includegraphics[width=0.3\textwidth]{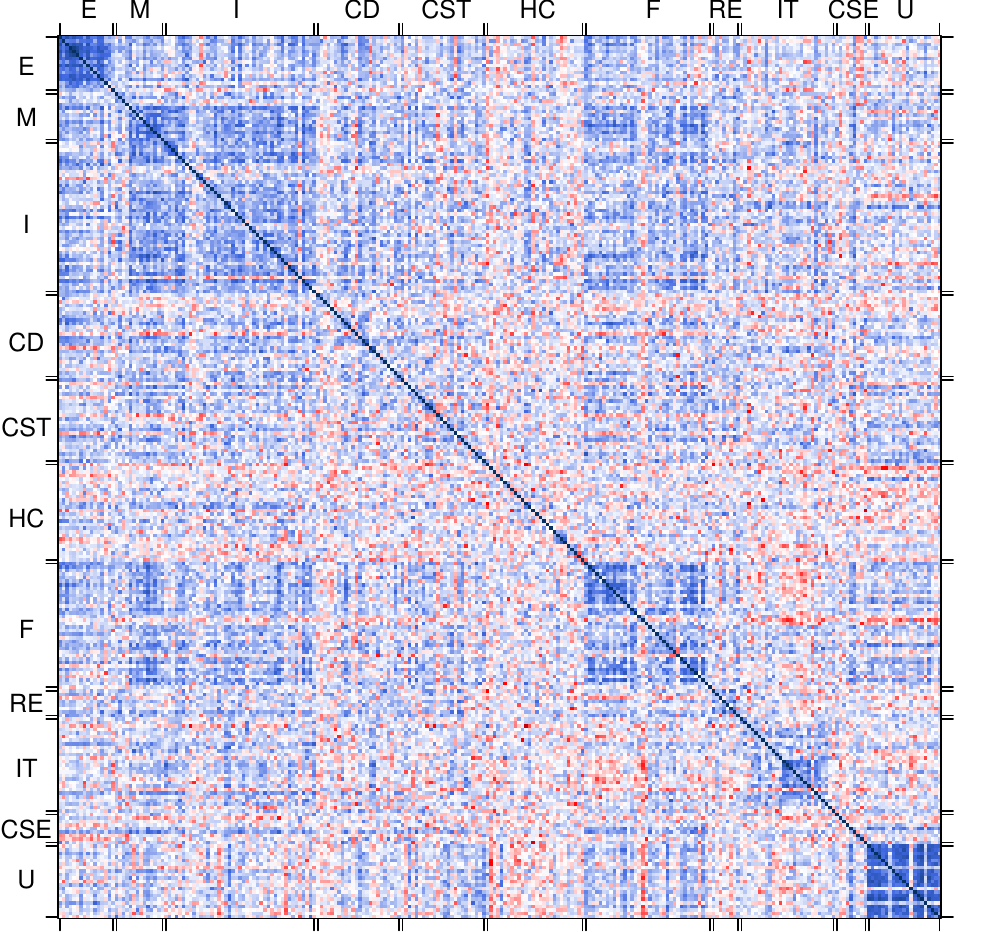}
		}
		\subfloat[\label{subfig:CovApprDC_2_Corr}2 epochs after (DC).]{
			\includegraphics[width=0.3\textwidth]{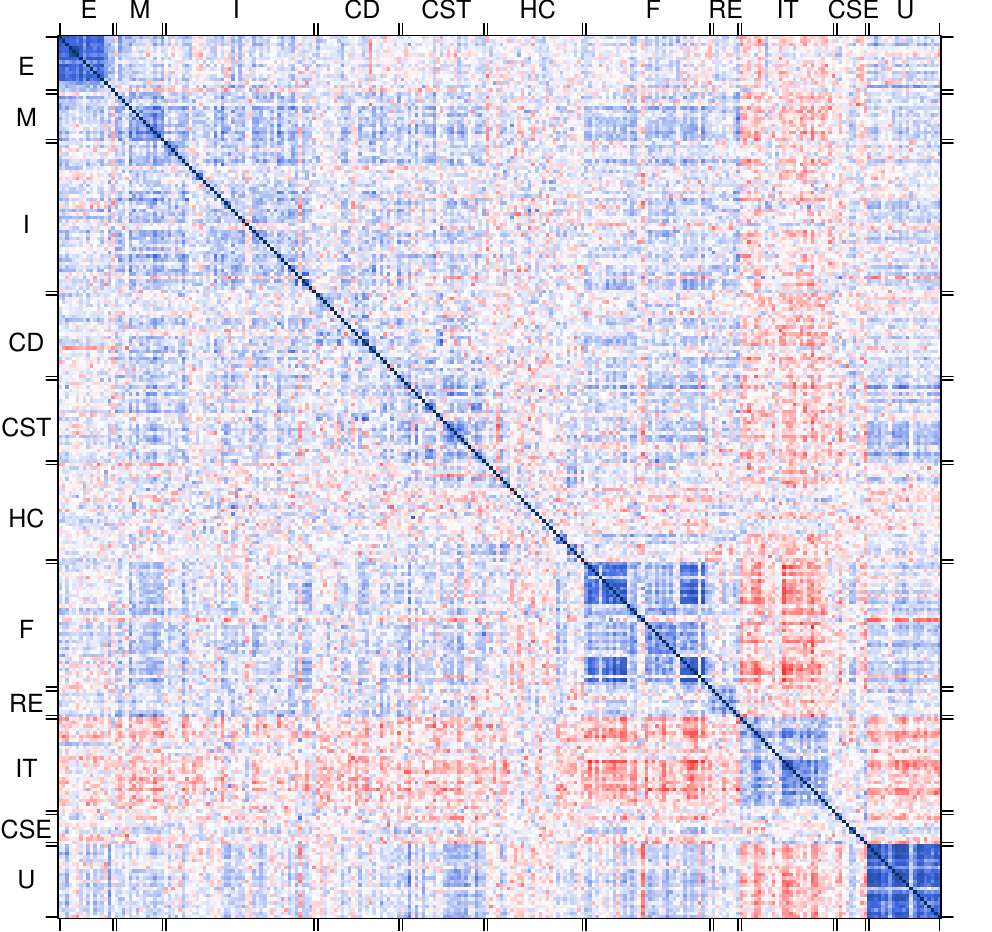}
		}
		\subfloat[\label{subfig:CovApprDC_4_Corr}4 epochs after (DC).]{
			\includegraphics[width=0.3\textwidth]{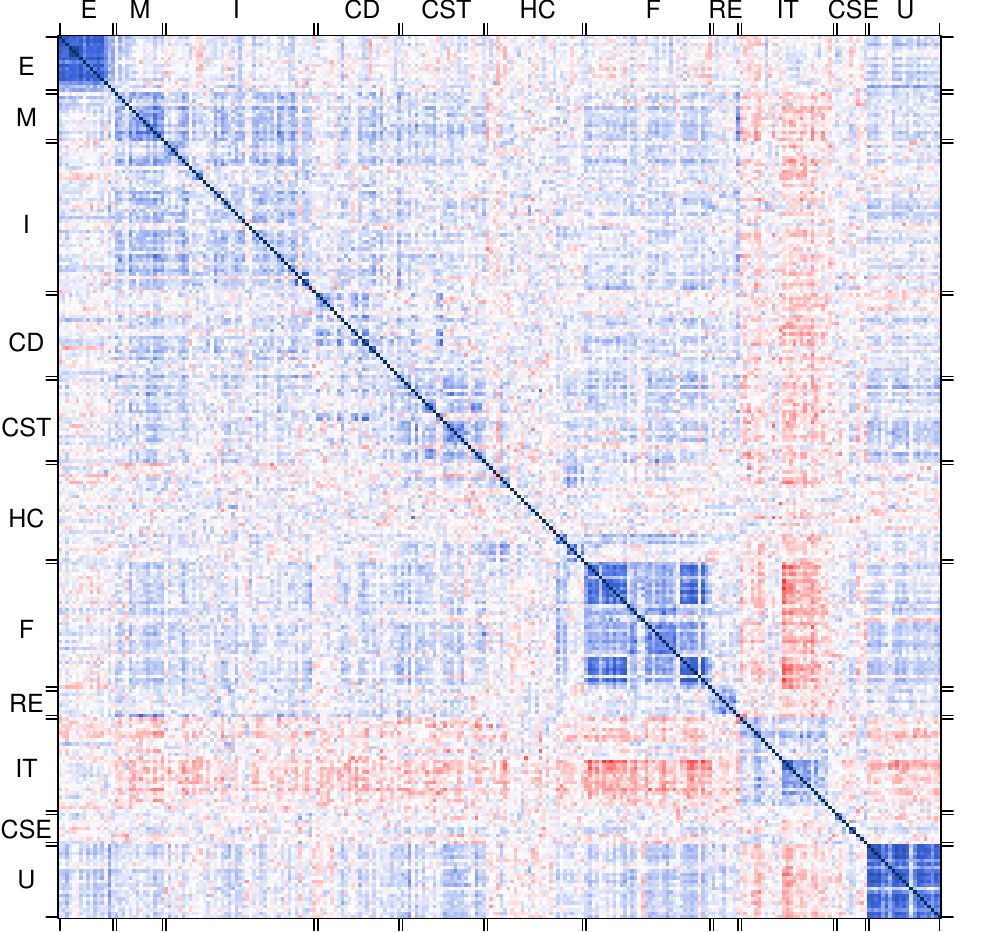}
		}
	\end{minipage}%
	\\
	\begin{minipage}{1.0\textwidth}
		\subfloat[\label{subfig:CorrApprDC_1_Corr}0 epochs after (DC).]{
			\includegraphics[width=0.3\textwidth]{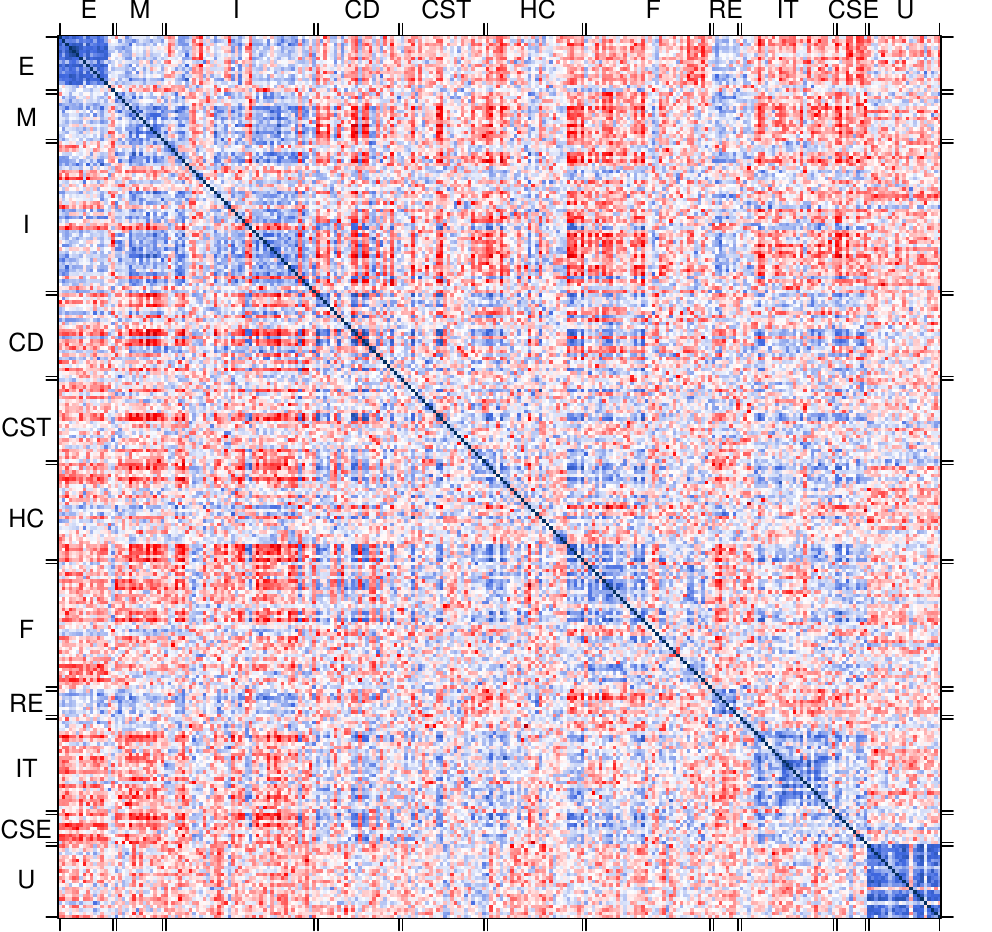}
		}
		\subfloat[\label{subfig:CorrApprDC_2_Corr}6 epochs after (DC).]{
			\includegraphics[width=0.3\textwidth]{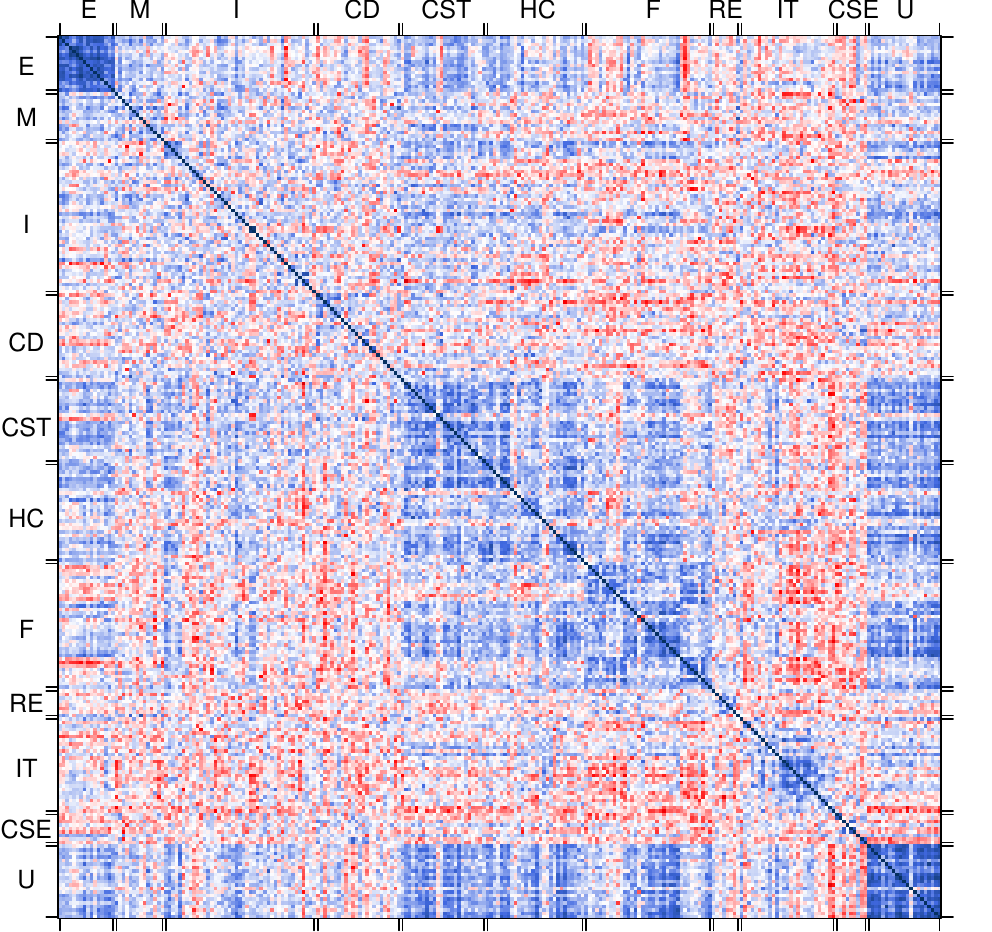}
		}
		\subfloat[\label{subfig:CorrApprDC_3_Corr}7 epochs after (DC).]{
			\includegraphics[width=0.3\textwidth]{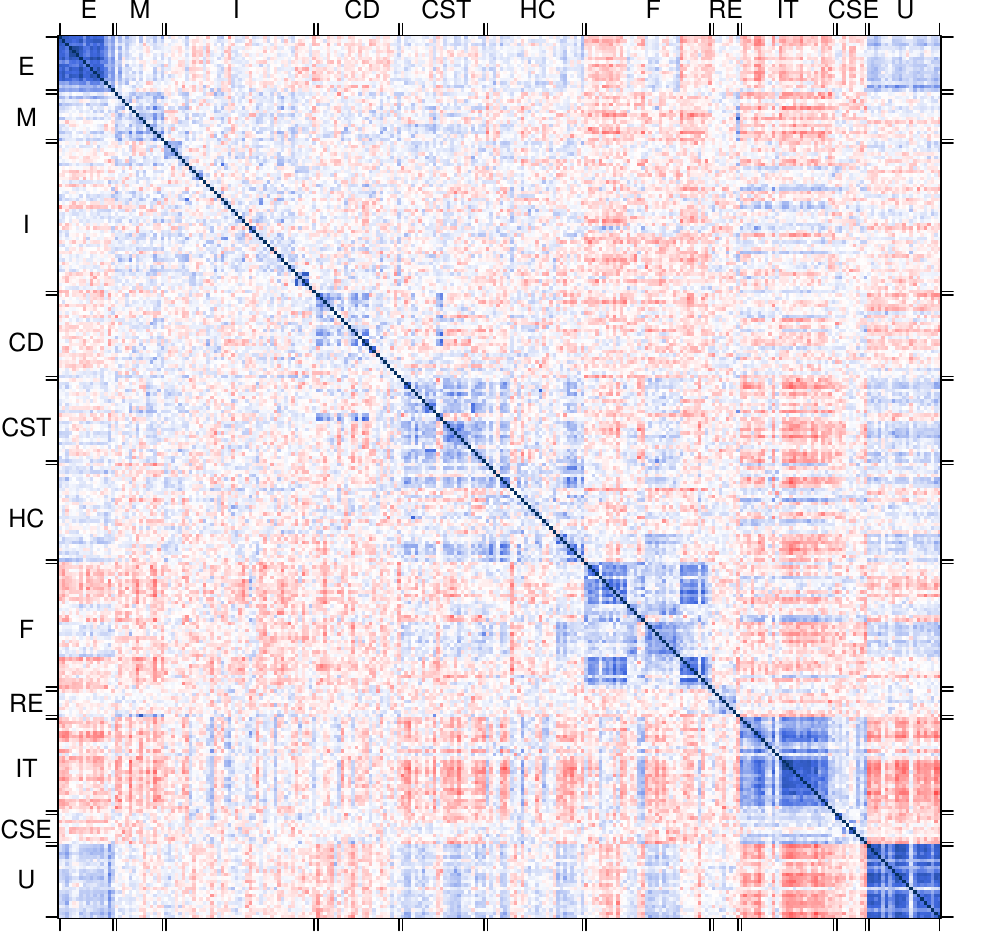}
		}
	\end{minipage}
	\caption{\label{subfig:Main:CovApprDC_Corr}Typical second market states ($K = 250$ stocks) corresponding to the snapshots of the dot-com bubble burst marked by historical event (DC) (cf.~Fig.~\ref{subfig:Main:CovApprLB}) for the reduced-rank correlation matrices in the \protect\subref{subfig:CovApprDC_1_Corr}-\protect\subref{subfig:CovApprDC_4_Corr} covariance approach and \protect\subref{subfig:CorrApprDC_1_Corr}-\protect\subref{subfig:CorrApprDC_3_Corr} correlation approach.
		A typical market state is obtained by taking the element-wise average of the correlation matrices of the respective market state. 
		From subplots \protect\subref{subfig:CovApprDC_1_Corr} to \protect\subref{subfig:CorrApprDC_3_Corr}, the number of the epochs after (DC) is specified. Capital Letters indicate industrial sectors (see~Tab.~\ref{tab:GICS})
		(\href{https://www.quandl.com/}{Data 
			from QuoteMedia via Quandl}).}
\end{figure*}

In Ref.~\cite{Heckens_2020}, we divided a 15 year time period into epochs (time periods of equal length) with $T_{\text{ep}} = 42$~trading days and set up a $K \times T_{\text{ep}}$~data matrix $A$ and $M$ with $K= 262$ and $T_{\text{ep}} = 42$ for each epoch. Then we calculated the reduced-rank correlation matrices for these epochs and clustered all reduced-rank correlation matrices by the $k$-means algorithm~\cite{Steinhaus1956:DivisionCorpMaterielEnParties,
	BallHall1965:ISODATA, MacQueen1967:SomeMethodsClassification,
	LLoyd1982:LeastSquaresQuantization, Jain_2010}, resulting in different market states.
According to Secs.~\ref{sec:CovAppr} and \ref{sec:MeanCaluesDistanceMatricesAndAverageDistances}, we predominantly cluster information on the industrial sectors.
The market state analysis reveals much more clearly major structures hidden in the distance matrices
as we strongly reduce the information of the reduced-rank correlation matrices to the market states.

Here, we follow a different method which we want to explain in the case of the Lehman Brothers crash (LB) by means of Figs.~\subref*{subfig:CovApprLB_1}--\subref*{subfig:CovApprLB_5} for the covariance approach.
We divide a time period of several years before the Lehman Brothers crash (LB) into epochs with $T_{\text{ep}} = 42$ trading days.
Figs.~\subref*{subfig:CovApprLB_1}--\subref*{subfig:CovApprLB_5} have in common that the starting epoch of the time period is fixed.
In Fig.~\subref*{subfig:CovApprLB_1}, the last epoch is seven epochs before (LB) and in Fig.~\subref*{subfig:CovApprLB_5} one epoch before (LB).
For each epoch, we calculate a reduced-rank correlation matrix of dimension $250 \times 250$ and cluster all reduced-rank correlation matrices employing the $k$-means clustering algorithm for $k=2$.
We refer to Figs.~\subref*{subfig:CovApprLB_1}--\subref*{subfig:CovApprLB_5} as ``snapshots''.
All snapshots together for one approach form a sequence of snapshots.

For the first snapshot in Fig.~\subref*{subfig:CovApprLB_1}, we divide the time period 
from \mbox{2004-01-15} to \mbox{2007-07-18} into 21 epochs. 
The last epoch is in the vicinity of where we observe a small value of
the mean value $\mean{\text{corr}}_B$
(see~Sec.~\ref{sec:OtherLongTermPrecursors}), \textit{i.e.} between (LB1) and (LB2) (see~Tab.~\ref{tab:EventsMarktStateTrans}).
In Fig.~\subref*{subfig:CovApprLB_2}, we divide the time period from 2004-01-15 to 2007-09-17 into 22 epochs. 
We cluster 22 reduced-rank correlation matrices, \textit{i.e.} additionally, we cluster one more reduced-rank correlation matrix corresponding to an epoch of 42 trading days.
In contrast to Fig.~\subref*{subfig:CovApprLB_1}, we observe in Fig.~\subref*{subfig:CovApprLB_2} a drastic change in the market state evolution.
For the first 21 epochs, all reduced-rank correlation matrices are clustered into the first market state. 
The second market state consists of one reduced-rank correlation matrix located between (LB1) and (LB2).
In Fig.~\subref*{subfig:CovApprLB_3}, we consider 25 epochs, \textit{i.e.} three more reduced-rank correlation matrices compared to Fig.~\subref*{subfig:CovApprLB_2}.
The two remaining reduced-rank correlation matrices in Figs.~\subref*{subfig:CovApprLB_4} and~\subref*{subfig:CovApprLB_5} are assigned to the second market state as well, revealing a certain stability in the market states.

The sequence of snapshots for the correlation approach in Figs.~\subref*{subfig:CorrApprLB_1}--\subref*{subfig:CorrApprLB_5} looks very similar compared to the one for the covariance approach in Figs.~\subref*{subfig:CovApprLB_1}--\subref*{subfig:CovApprLB_5}.
For both approaches, there are three epochs between the last epoch in Figs.~\subref*{subfig:CovApprLB_3} and~\subref*{subfig:CorrApprLB_3} and event (LB) which is half of a trading year.
We are able to detect such a structural change without using post-crash data, thereby demonstrating precursor properties of the reduced-rank correlation matrices. 

We obtain very similar results for the covariance and the correlation approach which is supported by the plots in Fig.~\ref{subfig:Main:CorrApprLB_Corr}.
Here, we show so-called \textit{typical} market states~\cite{munnixIdentifyingStatesFinancial2012,Heckens_2020} as element-wise average of all reduced-rank correlation matrices of a single market state
for the second market state and for different snapshots.
Corresponding typical second market states for the covariance and the correlation approach are
very similar.

For dot-com bubble burst in Fig.~\ref{subfig:Main:CovApprDC}, 
we are not able to identify a market state transition before (DC).
Major correlation structure changes appear afterwards.
The clustered period starts for all snapshots at 1997-01-13.
Estimated events~(DC1) and~(DC2) (see~Tab.~\ref{tab:EventsMarktStateTrans}) highlight the market state transition and small values of mean value $\mean{\text{corr}}_B$
(cf.~Sec.~\ref{sec:OtherLongTermPrecursors}).
Instead, we take into account epochs after (DC) and analyze for which snapshots the market state transitions can be detected.
In contrast to the Lehman Brothers crash, the temporal evolutions of the typical market states
within the snapshots for the covariance approach differ qualitatively from those of the correlation approach in the case of the dot-com bubble burst.
It is intriguing that we are able to detect a  first market state transition after two epochs in Fig.~\subref*{subfig:CovApprDC_2} which is much earlier than the six epochs after (DC) for the correlation approach in Fig.~\subref*{subfig:CorrApprDC_2}.

We aim to compare the mean correlation $\mean{\text{corr}}_B$ with a time series mimicking the NASDAQ index~\cite{wiki:2020:DeadCatBounceNASDAQ}.
We construct this time series by taking the average of the adjusted daily closing prices belonging to the 27 IT stocks (cf.~Fig.~\ref{tab:GICS}).
In Fig.~\ref{fig:IndexDC},
this self-constructed index is displayed (cf.~Tab.~\ref{tab:GICS} and Tab.~\ref{tab:OverviewSP500} in App.~\ref{sec:ListStocks}).
The end date of the second epoch after (DC) is 2000-07-11 and highlighted by the label (EP2);
the end date of the fourth epoch after (DC) is 2000-11-07 and highlighted by the label (EP4).
Event (EP2) lies before a major drop of
the dot-com bubble burst in Fig.~\ref{fig:IndexDC}. 
It is interesting to
compare this observation with the temporal evolution of the mean values in
Sec.~\ref{sec:MeanCaluesDistanceMatricesAndAverageDistances}.
In Fig.~\subref*{subfig:CovRedRankMean} the mean correlation $\mean{\text{corr}}_B$ remains at a relatively high level
whereas the mean covariance $\mean{\text{cov}}$ is decreasing in Fig.~\subref*{subfig:CovStandMean} and Fig.~\ref{fig:MeanCovStandLOG} on a logarithmic scale.
The mean correlation $\mean{\text{corr}}_B$ seems to indicate that there is still an endogenous risk in the IT sector which finally results in the market drop.
This is another result which adds to the observation of 
Sec.~\ref{sec:OtherLongTermPrecursors} that the mean correlation $\mean{\text{corr}}_B$ is a potential measure for systemic risk.
 
We can also observe in the correlation structure of the typical second market states in Fig.~\ref{subfig:Main:CovApprDC_Corr}
that both market state transitions are different for the two approaches.
The anti-correlations from the IT sector with the other ten sectors as in Fig.~\subref*{subfig:CovApprDC_4_Corr} also appear as a feature of the reduced-rank correlation matrices in the covariance approach around the Lehman Brothers crash according to Ref.~\cite{Heckens_2020}.
Based on this observation and the historical events during the time period, the second market state in the time period of the dot-com bubble burst can be referred to as ``crisis state" as well.

\section{\label{sec:Conclusion}Conclusion}

We studied the dynamics of reduced-rank correlation matrices in the covariance and the correlation approach
and found long-term precursor properties.
The dynamics of the correlation structure was analyzed in two different ways. On the one hand we looked at the market states of the reduced-rank correlation matrices, on the other hand we were able to relate the market state transitions to sometimes large peaks or drastic changes in the 
mean values of covariance and correlation approach and in the corresponding averaged distances.

Analyzing the variance of the largest four eigenvalues for standard and reduced-
rank matrices we corroborated the quasi-stationary behavior of the market states based on the reduced-rank correlation matrices from Ref.~\cite{Heckens_2020}. Moreover, we found that the industrial sectors show a coherent behavior over the entire time period.

We have introduced a new method for analyzing market states. With our technique of snapshots for the market state analysis, we can follow the market state transitions by adding reduced-rank correlation matrices of new epochs to the market state analysis, thereby following the trajectories of the reduced-rank correlation matrices before or within crises periods.
Our market state analysis is able to detect market state transitions belonging to the Lehman Brothers crash occurred.
We exclusively used pre-crises data. 
The snapshots for the covariance and correlation approach look very much alike.
For the dot-com bubble burst both approaches reveal differences concerning the market state dynamics.

We identified
structures in the averaged distances which coincide with market state transitions in our cluster analysis.
The mid-2007 event (freezing of the Interbank market) is the very first precursor signal starting the Lehman Brothers crisis state.
The burst of the dot-com bubble marks the beginning of a crisis state as well.

The market state transition prior to the Lehman crash occurred at the time of the Interbank lending freeze. In the averaged distances, the influence of this event dampened.
Nevertheless, we still detect the second market state shortly before the Lehman crash.
Hence we observe a hysteresis effect, \textit{i.e.} the second market state shortly before the Lehman crash was not directly caused by the Interbank lending freeze. Nonetheless the market stayed in the crisis market state meaning that the market was still in a period not necessarily but potentially leading to a crash.

By comparing the mean correlations in the covariance and correlation approach with historical events we saw that the mean correlations (especially in the covariance approach)
reflect the dynamics of an endogenous risks
for both crises.
Points of minimum correlations in the covariance approach
indicate the beginning of two crises periods.
From these points onwards, the endogenous risk builds up. Thus it is
conceivable that a market state transition takes place since a new economic period begins in terms of
the endogenous risk dynamics.

Our new market state method relates precursors of different kinds.
In the market state analysis, transitions into a crises market state are
connected to low mean correlations in the covariance approach and sudden changes in the averaged distances.

Furthermore, we found a period around the Lehman Brothers crash coinciding with a recession. Usually, recession periods are calculated with the gross domestic product (GDP) and techniques such as the Hodrick-Prescott filter \cite{hodrick1997postwar,Hamilton_2018,Hodrick_2020}. Analogously, by subtracting the dyadic matrix corresponding to the largest eigenvalue, we also separated the quickly changing market motion from the more stable sectoral one.

For the Lehman Brothers crisis state,
changes in the averaged distances build up prior to the changes in the mean correlation at a market state transition.
The interpretation for this phenomenon might be that economic changes like the freezing of the Interbank market~\cite{Investopedia}
are first visible which is supported by our identification of a ``recession market state". 
The endogenous risk increases as some kind of ``economic tension" in the market which builds up and finally dissipates. This risk is potentially contagious for the entire market. In the case of the Lehman Brothers crash
it might be viewed as a spill-over effect.
All our observations lead to the conclusion that the mean correlations for the reduced-rank correlation matrices in both approaches describe to some extent the fragility of the market being exposed to a potentially larger risk initialized by an industrial sector. In the covariance approach, we observed anti-correlations between the IT-sector and the other ten sectors during the dot-com bubble burst. Such anti-correlations are also visible in the Lehman Brothers crash pre-phase between the financial sector and the other ten sectors in Ref.~\cite{Heckens_2020}.
Therefore, the above mentioned endogenous risks, visible in the mean correlations in the covariance and correlation approach, are potential measures for systemic risk.

\clearpage
\bibliography{Lit.bib}

\appendix

\clearpage
\onecolumngrid
\section{\label{sec:ListStocks}List of selected stocks}

{\tiny
	\begin{longtable}{rrllp{4cm}}
		\caption[]{Overview of the 250 selected stocks of the S\&P 500 index (cf.~{\cite{wiki:2020:LISTSP500}}).} 

		\label{tab:OverviewSP500} \\
		\toprule
		Number & Symbol & Security & Sector & Sub-Industry\\
		\midrule\endfirsthead
		\caption*{Continuation: Overview of the 250 selected stocks of the S\&P 500 
index (cf.~{\cite{wiki:2020:LISTSP500}}).}  \\
		\toprule
		Number & Symbol & Security & Sector & Sub-Industry\\
		\midrule\endhead
1&CVX&Chevron Corp.&Energy&Integrated Oil \& Gas \\
2&HES&Hess Corporation&Energy&Integrated Oil \& Gas \\
3&XOM&Exxon Mobil Corp.&Energy&Integrated Oil \& Gas \\
4&BKR&Baker Hughes Co&Energy&Oil \& Gas Equipment \& Services \\
5&HAL&Halliburton Co.&Energy&Oil \& Gas Equipment \& Services \\
6&SLB&Schlumberger Ltd.&Energy&Oil \& Gas Equipment \& Services \\
7&APA&Apache Corporation&Energy&Oil \& Gas Exploration \& Production \\
8&COG&Cabot Oil \& Gas&Energy&Oil \& Gas Exploration \& Production \\
9&COP&ConocoPhillips&Energy&Oil \& Gas Exploration \& Production \\
10&EOG&EOG Resources&Energy&Oil \& Gas Exploration \& Production \\
11&MRO&Marathon Oil Corp.&Energy&Oil \& Gas Exploration \& Production \\
12&NBL&Noble Energy&Energy&Oil \& Gas Exploration \& Production \\
13&OXY&Occidental Petroleum&Energy&Oil \& Gas Exploration \& Production \\
14&VLO&Valero Energy&Energy&Oil \& Gas Refining \& Marketing \\
15&OKE&ONEOK&Energy&Oil \& Gas Storage \& Transportation \\
16&WMB&Williams Companies&Energy&Oil \& Gas Storage \& Transportation \\
17&VMC&Vulcan Materials&Materials&Construction Materials \\
18&FMC&FMC Corporation&Materials&Fertilizers \& Agricultural Chemicals \\
19&MOS&The Mosaic Company&Materials&Fertilizers \& Agricultural Chemicals \\
20&NEM&Newmont Corporation&Materials&Gold \\
21&APD&Air Products \& Chemicals Inc&Materials&Industrial Gases \\
22&BLL&Ball Corp&Materials&Metal \& Glass Containers \\
23&AVY&Avery Dennison Corp&Materials&Paper Packaging \\
24&IP&International Paper&Materials&Paper Packaging \\
25&SEE&Sealed Air&Materials&Paper Packaging \\
26&ECL&Ecolab Inc.&Materials&Specialty Chemicals \\
27&IFF&International Flavors \& Fragrances&Materials&Specialty Chemicals \\
28&PPG&PPG Industries&Materials&Specialty Chemicals \\
29&SHW&Sherwin-Williams&Materials&Specialty Chemicals \\
30&NUE&Nucor Corp.&Materials&Steel \\
31&BA&Boeing Company&Industrials&Aerospace \& Defense \\
32&GD&General Dynamics&Industrials&Aerospace \& Defense \\
33&LMT&Lockheed Martin Corp.&Industrials&Aerospace \& Defense \\
34&NOC&Northrop Grumman&Industrials&Aerospace \& Defense \\
35&RTX&Raytheon Technologies&Industrials&Aerospace \& Defense \\
36&TXT&Textron Inc.&Industrials&Aerospace \& Defense \\
37&DE&Deere \& Co.&Industrials&Agricultural \& Farm Machinery \\
38&EXPD&Expeditors&Industrials&Air Freight \& Logistics \\
39&FDX&FedEx Corporation&Industrials&Air Freight \& Logistics \\
40&ALK&Alaska Air Group Inc&Industrials&Airlines \\
41&LUV&Southwest Airlines&Industrials&Airlines \\
42&AOS&A.O. Smith Corp&Industrials&Building Products \\
43&FAST&Fastenal Co&Industrials&Building Products \\
44&JCI&Johnson Controls International&Industrials&Building Products \\
45&MAS&Masco Corp.&Industrials&Building Products \\
46&J&Jacobs Engineering Group&Industrials&Construction \& Engineering \\
47&CAT&Caterpillar Inc.&Industrials&Construction Machinery \& Heavy Trucks \\
48&PCAR&PACCAR Inc.&Industrials&Construction Machinery \& Heavy Trucks \\
49&CTAS&Cintas Corporation&Industrials&Diversified Support Services \\
50&AME&AMETEK Inc.&Industrials&Electrical Components \& Equipment \\
51&EMR&Emerson Electric Company&Industrials&Electrical Components \& Equipment \\
52&ETN&Eaton Corporation&Industrials&Electrical Components \& Equipment \\
53&ROK&Rockwell Automation Inc.&Industrials&Electrical Components \& Equipment \\
54&ROL&Rollins Inc.&Industrials&Environmental \& Facilities Services \\
55&GE&General Electric&Industrials&Industrial Conglomerates \\
56&HON&Honeywell Int'l Inc.&Industrials&Industrial Conglomerates \\
57&MMM&3M Company&Industrials&Industrial Conglomerates \\
58&CMI&Cummins Inc.&Industrials&Industrial Machinery \\
59&DOV&Dover Corporation&Industrials&Industrial Machinery \\
60&FLS&Flowserve Corporation&Industrials&Industrial Machinery \\
61&GWW&Grainger (W.W.) Inc.&Industrials&Industrial Machinery \\
62&IEX&IDEX Corporation&Industrials&Industrial Machinery \\
63&ITW&Illinois Tool Works&Industrials&Industrial Machinery \\
64&PH&Parker-Hannifin&Industrials&Industrial Machinery \\
65&PNR&Pentair plc&Industrials&Industrial Machinery \\
66&SNA&Snap-on&Industrials&Industrial Machinery \\
67&SWK&Stanley Black \& Decker&Industrials&Industrial Machinery \\
68&CSX&CSX Corp.&Industrials&Railroads \\
69&KSU&Kansas City Southern&Industrials&Railroads \\
70&NSC&Norfolk Southern Corp.&Industrials&Railroads \\
71&UNP&Union Pacific Corp&Industrials&Railroads \\
72&EFX&Equifax Inc.&Industrials&Research \& Consulting Services \\
73&JBHT&J. B. Hunt Transport Services&Industrials&Trucking \\
74&GPS&Gap Inc.&Consumer Discretionary&Apparel Retail \\
75&LB&L Brands Inc.&Consumer Discretionary&Apparel Retail \\
76&ROST&Ross Stores&Consumer Discretionary&Apparel Retail \\
77&TJX&TJX Companies Inc.&Consumer Discretionary&Apparel Retail \\
78&NKE&Nike, Inc.&Consumer Discretionary&Apparel, Accessories \& Luxury Goods \\
79&PVH&PVH Corp.&Consumer Discretionary&Apparel, Accessories \& Luxury Goods \\
80&TIF&Tiffany \& Co.&Consumer Discretionary&Apparel, Accessories \& Luxury Goods \\
81&VFC&VF Corporation&Consumer Discretionary&Apparel, Accessories \& Luxury Goods \\
82&F&Ford Motor Company&Consumer Discretionary&Automobile Manufacturers \\
83&MGM&MGM Resorts International&Consumer Discretionary&Casinos \& Gaming \\
84&BBY&Best Buy Co. Inc.&Consumer Discretionary&Computer \& Electronics Retail \\
85&TGT&Target Corp.&Consumer Discretionary&General Merchandise Stores \\
86&LEG&Leggett \& Platt&Consumer Discretionary&Home Furnishings \\
87&HD&Home Depot&Consumer Discretionary&Home Improvement Retail \\
88&LOW&Lowe's Cos.&Consumer Discretionary&Home Improvement Retail \\
89&LEN&Lennar Corp.&Consumer Discretionary&Homebuilding \\
90&PHM&PulteGroup&Consumer Discretionary&Homebuilding \\
91&CCL&Carnival Corp.&Consumer Discretionary&Hotels, Resorts \& Cruise Lines \\
92&WHR&Whirlpool Corp.&Consumer Discretionary&Household Appliances \\
93&NWL&Newell Brands&Consumer Discretionary&Housewares \& Specialties \\
94&HAS&Hasbro Inc.&Consumer Discretionary&Leisure Products \\
95&MCD&McDonald's Corp.&Consumer Discretionary&Restaurants \\
96&HRB&H\&R Block&Consumer Discretionary&Specialized Consumer Services \\
97&GPC&Genuine Parts&Consumer Discretionary&Specialty Stores \\
98&ADM&Archer-Daniels-Midland Co&Consumer Staples&Agricultural Products \\
99&TAP&Molson Coors Beverage Company&Consumer Staples&Brewers \\
100&BF.B&Brown-Forman Corp.&Consumer Staples&Distillers \& Vintners \\
101&WBA&Walgreens Boots Alliance&Consumer Staples&Drug Retail \\
102&SYY&Sysco Corp.&Consumer Staples&Food Distributors \\
103&KR&Kroger Co.&Consumer Staples&Food Retail \\
104&CHD&Church \& Dwight&Consumer Staples&Household Products \\
105&CL&Colgate-Palmolive&Consumer Staples&Household Products \\
106&CLX&The Clorox Company&Consumer Staples&Household Products \\
107&KMB&Kimberly-Clark&Consumer Staples&Household Products \\
108&COST&Costco Wholesale Corp.&Consumer Staples&Hypermarkets \& Super Centers \\
109&WMT&Walmart&Consumer Staples&Hypermarkets \& Super Centers \\
110&CAG&Conagra Brands&Consumer Staples&Packaged Foods \& Meats \\
111&CPB&Campbell Soup&Consumer Staples&Packaged Foods \& Meats \\
112&GIS&General Mills&Consumer Staples&Packaged Foods \& Meats \\
113&HRL&Hormel Foods Corp.&Consumer Staples&Packaged Foods \& Meats \\
114&HSY&The Hershey Company&Consumer Staples&Packaged Foods \& Meats \\
115&K&Kellogg Co.&Consumer Staples&Packaged Foods \& Meats \\
116&MKC&McCormick \& Co.&Consumer Staples&Packaged Foods \& Meats \\
117&TSN&Tyson Foods&Consumer Staples&Packaged Foods \& Meats \\
118&PG&Procter \& Gamble&Consumer Staples&Personal Products \\
119&KO&Coca-Cola Company&Consumer Staples&Soft Drinks \\
120&PEP&PepsiCo Inc.&Consumer Staples&Soft Drinks \\
121&MO&Altria Group Inc&Consumer Staples&Tobacco \\
122&AMGN&Amgen Inc.&Health Care&Biotechnology \\
123&BMY&Bristol-Myers Squibb&Health Care&Health Care Distributors \\
124&CAH&Cardinal Health Inc.&Health Care&Health Care Distributors \\
125&ABMD&ABIOMED Inc&Health Care&Health Care Equipment \\
126&ABT&Abbott Laboratories&Health Care&Health Care Equipment \\
127&BAX&Baxter International Inc.&Health Care&Health Care Equipment \\
128&BDX&Becton Dickinson&Health Care&Health Care Equipment \\
129&DHR&Danaher Corp.&Health Care&Health Care Equipment \\
130&HOLX&Hologic&Health Care&Health Care Equipment \\
131&MDT&Medtronic plc&Health Care&Health Care Equipment \\
132&PKI&PerkinElmer&Health Care&Health Care Equipment \\
133&SYK&Stryker Corp.&Health Care&Health Care Equipment \\
134&TFX&Teleflex&Health Care&Health Care Equipment \\
135&VAR&Varian Medical Systems&Health Care&Health Care Equipment \\
136&UHS&Universal Health Services&Health Care&Health Care Facilities \\
137&CVS&CVS Health&Health Care&Health Care Services \\
138&WST&West Pharmaceutical Services&Health Care&Health Care Supplies \\
139&CERN&Cerner&Health Care&Health Care Technology \\
140&BIO&Bio-Rad Laboratories&Health Care&Life Sciences Tools \& Services \\
141&TMO&Thermo Fisher Scientific&Health Care&Life Sciences Tools \& Services \\
142&CI&CIGNA Corp.&Health Care&Managed Health Care \\
143&HUM&Humana Inc.&Health Care&Managed Health Care \\
144&UNH&United Health Group Inc.&Health Care&Managed Health Care \\
145&JNJ&Johnson \& Johnson&Health Care&Pharmaceuticals \\
146&LLY&Lilly (Eli) \& Co.&Health Care&Pharmaceuticals \\
147&MRK&Merck \& Co.&Health Care&Pharmaceuticals \\
148&MYL&Mylan N.V.&Health Care&Pharmaceuticals \\
149&PFE&Pfizer Inc.&Health Care&Pharmaceuticals \\
150&BEN&Franklin Resources&Financials&Asset Management \& Custody Banks \\
151&BK&The Bank of New York Mellon&Financials&Asset Management \& Custody Banks \\
152&NTRS&Northern Trust Corp.&Financials&Asset Management \& Custody Banks \\
153&STT&State Street Corp.&Financials&Asset Management \& Custody Banks \\
154&TROW&T. Rowe Price Group&Financials&Asset Management \& Custody Banks \\
155&AXP&American Express Co&Financials&Consumer Finance \\
156&BAC&Bank of America Corp&Financials&Diversified Banks \\
157&C&Citigroup Inc.&Financials&Diversified Banks \\
158&CMA&Comerica Inc.&Financials&Diversified Banks \\
159&JPM&JPMorgan Chase \& Co.&Financials&Diversified Banks \\
160&USB&U.S. Bancorp&Financials&Diversified Banks \\
161&WFC&Wells Fargo&Financials&Diversified Banks \\
162&AJG&Arthur J. Gallagher \& Co.&Financials&Insurance Brokers \\
163&AON&Aon plc&Financials&Insurance Brokers \\
164&MMC&Marsh \& McLennan&Financials&Insurance Brokers \\
165&RJF&Raymond James Financial Inc.&Financials&Investment Banking \& Brokerage \\
166&SCHW&Charles Schwab Corporation&Financials&Investment Banking \& Brokerage \\
167&AFL&AFLAC Inc&Financials&Life \& Health Insurance \\
168&GL&Globe Life Inc.&Financials&Life \& Health Insurance \\
169&UNM&Unum Group&Financials&Life \& Health Insurance \\
170&L&Loews Corp.&Financials&Multi-line Insurance \\
171&LNC&Lincoln National&Financials&Multi-line Insurance \\
172&AIG&American International Group&Financials&Property \& Casualty Insurance \\
173&CINF&Cincinnati Financial&Financials&Property \& Casualty Insurance \\
174&PGR&Progressive Corp.&Financials&Property \& Casualty Insurance \\
175&TRV&The Travelers Companies Inc.&Financials&Property \& Casualty Insurance \\
176&WRB&W. R. Berkley Corporation&Financials&Property \& Casualty Insurance \\
177&FITB&Fifth Third Bancorp&Financials&Regional Banks \\
178&HBAN&Huntington Bancshares&Financials&Regional Banks \\
179&KEY&KeyCorp&Financials&Regional Banks \\
180&PNC&PNC Financial Services&Financials&Regional Banks \\
181&RF&Regions Financial Corp.&Financials&Regional Banks \\
182&SIVB&SVB Financial&Financials&Regional Banks \\
183&TFC&Truist Financial&Financials&Regional Banks \\
184&ZION&Zions Bancorp&Financials&Regional Banks \\
185&PBCT&People's United Financial&Financials&Thrifts \& Mortgage Finance \\
186&PEAK&Healthpeak Properties&Real Estate&Health Care REITs \\
187&HST&Host Hotels \& Resorts&Real Estate&Hotel \& Resort REITs \\
188&DRE&Duke Realty Corp&Real Estate&Industrial REITs \\
189&VNO&Vornado Realty Trust&Real Estate&Office REITs \\
190&UDR&UDR, Inc.&Real Estate&Residential REITs \\
191&FRT&Federal Realty Investment Trust&Real Estate&Retail REITs \\
192&PSA&Public Storage&Real Estate&Specialized REITs \\
193&WY&Weyerhaeuser&Real Estate&Specialized REITs \\
194&ADBE&Adobe Inc.&Information Technology&Application Software \\
195&ADSK&Autodesk Inc.&Information Technology&Application Software \\
196&CDNS&Cadence Design Systems&Information Technology&Application Software \\
197&NLOK&NortonLifeLock&Information Technology&Application Software \\
198&ORCL&Oracle Corp.&Information Technology&Application Software \\
199&CSCO&Cisco Systems&Information Technology&Communications Equipment \\
200&MSI&Motorola Solutions Inc.&Information Technology&Communications Equipment \\
201&FISV&Fiserv Inc&Information Technology&Data Processing \& Outsourced Services \\
202&PAYX&Paychex Inc.&Information Technology&Data Processing \& Outsourced Services \\
203&GLW&Corning Inc.&Information Technology&Electronic Components \\
204&ADP&Automatic Data Processing&Information Technology&Internet Services \& Infrastructure \\
205&IBM&International Business Machines&Information Technology&IT Consulting \& Other Services \\
206&AMAT&Applied Materials Inc.&Information Technology&Semiconductor Equipment \\
207&KLAC&KLA Corporation&Information Technology&Semiconductor Equipment \\
208&LRCX&Lam Research&Information Technology&Semiconductor Equipment \\
209&ADI&Analog Devices, Inc.&Information Technology&Semiconductors \\
210&AMD&Advanced Micro Devices Inc&Information Technology&Semiconductors \\
211&INTC&Intel Corp.&Information Technology&Semiconductors \\
212&MU&Micron Technology&Information Technology&Semiconductors \\
213&MXIM&Maxim Integrated Products Inc&Information Technology&Semiconductors \\
214&SWKS&Skyworks Solutions&Information Technology&Semiconductors \\
215&TXN&Texas Instruments&Information Technology&Semiconductors \\
216&MSFT&Microsoft Corp.&Information Technology&Systems Software \\
217&AAPL&Apple Inc.&Information Technology&Technology Hardware, Storage \& Peripherals \\
218&HPQ&HP Inc.&Information Technology&Technology Hardware, Storage \& Peripherals \\
219&WDC&Western Digital&Information Technology&Technology Hardware, Storage \& Peripherals \\
220&XRX&Xerox&Information Technology&Technology Hardware, Storage \& Peripherals \\
221&IPG&Interpublic Group&Communication Services&Advertising \\
222&OMC&Omnicom Group&Communication Services&Advertising \\
223&CTL&CenturyLink Inc&Communication Services&Alternative Carriers \\
224&CMCSA&Comcast Corp.&Communication Services&Cable \& Satellite \\
225&T&AT\&T Inc.&Communication Services&Integrated Telecommunication Services \\
226&VZ&Verizon Communications&Communication Services&Integrated Telecommunication Services \\
227&EA&Electronic Arts&Communication Services&Interactive Home Entertainment \\
228&DIS&The Walt Disney Company&Communication Services&Movies \& Entertainment \\
229&FOX&Fox Corporation (Class B)&Communication Services&Movies \& Entertainment \\
230&AEP&American Electric Power&Utilities&Electric Utilities \\
231&D&Dominion Energy&Utilities&Electric Utilities \\
232&DUK&Duke Energy&Utilities&Electric Utilities \\
233&ED&Consolidated Edison&Utilities&Electric Utilities \\
234&EIX&Edison Int'l&Utilities&Electric Utilities \\
235&ETR&Entergy Corp.&Utilities&Electric Utilities \\
236&EVRG&Evergy&Utilities&Electric Utilities \\
237&LNT&Alliant Energy Corp&Utilities&Electric Utilities \\
238&PEG&Public Service Enterprise Group (PSEG)&Utilities&Electric Utilities \\
239&PPL&PPL Corp.&Utilities&Electric Utilities \\
240&SO&Southern Company&Utilities&Electric Utilities \\
241&WEC&WEC Energy Group&Utilities&Electric Utilities \\
242&ATO&Atmos Energy&Utilities&Gas Utilities \\
243&CMS&CMS Energy&Utilities&Multi-Utilities \\
244&CNP&CenterPoint Energy&Utilities&Multi-Utilities \\
245&DTE&DTE Energy Co.&Utilities&Multi-Utilities \\
246&EXC&Exelon Corp.&Utilities&Multi-Utilities \\
247&NEE&NextEra Energy&Utilities&Multi-Utilities \\
248&NI&NiSource Inc.&Utilities&Multi-Utilities \\
249&PNW&Pinnacle West Capital&Utilities&Multi-Utilities \\
250&XEL&Xcel Energy Inc&Utilities&Multi-Utilities \\
		\bottomrule
	\end{longtable}
}

\end{document}